\documentclass[11pt]{amsart}

\usepackage{graphicx}
\usepackage{float}
\usepackage{dcolumn}
\usepackage{bm}

\usepackage[utf8]{inputenc}
\usepackage[T1]{fontenc}
\usepackage{multirow}
\usepackage{mathptmx}

 \newtheorem{remark}{Remark}
 
\newcommand{\tb}{\widetilde{\beta}}
\newcommand{\tg}{\widetilde{\gamma}}

\title[Mathematical and Computer Modeling  of COVID-19 Transmission Dynamics in Bulgaria]
{Mathematical and Computer Modeling  of COVID-19 Transmission Dynamics in
Bulgaria by Time-depended Inverse SEIR Model}
\author[S. Margenov, N. Popivanov, I. Ugrinova, S. Harizanov, T. Hristov]
{Svetozar Margenov \and Nedyu Popivanov \and Iva Ugrinova \and Stanislav Harizanov \and Tsvetan Hristov}
\address{Institute of Information and Communication Technologies, Bulgarian Academy of 
Sciences, Acad. G. Bonchev, bl. 25A, 1113 Sofia, Bulgaria (margenov@parallel.bas.bg)}
\address{Institute of Information and Communication Technologies, Bulgarian
Academy of Sciences, Acad. G. Bonchev, bl. 25A, 1113 Sofia, Bulgaria and Faculty of Mathematics and Informatics,
Sofia University ''St. Kliment Ohridski'', 5 James Bourchier blvd., 1164 Sofia, Bulgaria (nedyu@parallel.bas.bg)}
\address{Institute of Molecular Biology, Bulgarian Academy of Sciences, Acad. G. Bonchev, bl. 21, 1113 Sofia, Bulgaria (ugryiva@gmail.com)}
\address{Institute of Information and Communication Technologies, Bulgarian Academy of 
Sciences, Acad. G. Bonchev, bl. 25A, 1113 Sofia, Bulgaria (sharizanov@parallel.bas.bg)}
\address{Faculty of Mathematics and Informatics,
Sofia University ''St. Kliment Ohridski'', 5 James Bourchier blvd., 1164 Sofia, Bulgaria (tsvetan@fmi.uni-sofia.bg)}

\begin{document}
\date{\today}

\begin{abstract} 

Since the end of 2019, with the outbreak of the new virus COVID-19, the world changed entirely in many aspects, with the pandemia affecting the economies, healthcare systems and the global socium. As a result from this pandemic, scientists from many countries across the globe united in their efforts to study the virsus's behavior and are attempting to predict mathematically its infection model in order to limit its impact and developing new methods and models to achieve this goal.
In this paper we explore a time-depended SEIR model, in which the
dynamics of the infection in four groups from a selected target
group (population), divided according to the infection, are modeled
by a system of  nonlinear ordinary differential equations.
 Several basic parameters are involved in the model: coefficients of infection rate,
 incubation rate, recovery rate. The coefficients are adaptable to each specific infection,
 for each individual country, and depend on the measures to limit the spread of
 the infection and the effectiveness of the methods of treatment of the infected
 people in the respective country. If such coefficients are known, solving the
 nonlinear system is possible to be able to make some hypotheses for the
 development of the epidemic. This is the reason for using Bulgarian COVID-19
 data to first of all, solve the so-called ''inverse problem'' and to find the parameters of the current situation.
Reverse logic is initially used to determine the parameters of the
model as a function of time, followed by computer solution of the
problem. Namely, this means predicting the future behavior of these
parameters, and finding  (and as a consequence applying mass-scale
measures, e.g., distancing, disinfection, limitation of public
events), a suitable scenario for  the change in the proportion of
the numbers of the four studied groups in the future. In fact, based
on these results we model the COVID-19 transmission dynamics in
Bulgaria and make a two-week forecast for the numbers of new cases
per day, active cases and recovered individuals. Such model, as we
show, has been successful for prediction analysis in the Bulgarian
situation. We also provide multiple examples of numerical
experiments with visualization of the results.
\end{abstract}

\maketitle

\section{Introduction}

\subsection{The Pandemic of SARS-CoV-2 - worldwide and in Bulgaria}

On 31st of December 2019, the Chinese Centre for Disease Control and
Prevention (China CDC) informed the World Health Organization (WHO)
about a cluster of 41 severe pneumonia cases of unknown aetiology in
Wuhan, Hubei province. A newly identified $\beta$-coronavirus (whole
genome sequenced on 5th of January and isolated on 7th of January)
 was initially named 2019-nCoV (2019-novel CoronaVirus/12 January 2020) and later
 renamed to SARS-CoV-2 (Severe Acute Respiratory Syndrome CoronaVirus -2/12 February
 2020) by the Coronavirus Study Group of the International Committee on Taxonomy of
  Viruses \cite{Touma}. The extensive global transmission led to the declaring
  coronavirus
  disease of 2019 (COVID-19) as a pandemic with significant mortality and morbidity \cite{Zhou}.
   By 27 February 2020, the outbreak of coronavirus disease 2019 (COVID-19) caused
   13 285 640 confirmed cases and 578 110 deaths globally, more than severe acute
   respiratory syndrome (SARS) (8273 cases, 775 deaths) and Middle East respiratory
   syndrome (MERS) (1139 cases, 431 deaths) caused in 2003 and 2013, respectively.
   COVID-19 has spread to 216 countries internationally. Total fatality rate of
   COVID-19 is estimated at 3.46\% by far based on published data from China CDC. Average incubation period
   of COVID-19 is around 6.4 days, ranges from 0 to 14 days. The basic reproductive
   number ($\Re_0$) of COVID-19 ranges from 2 to 3.5 at the early phase regardless of
   different prediction models, which is higher than SARS and MERS \cite{Wang}.
   The epidemiological and clinical data collected so far suggest that the disease
   spectrum of COVID-19 may differ from SARS or MERS. Person-to-person transmission
   of COVID-19 infection led to the isolation of patients that were subsequently
   administered a variety of treatments. Extensive measures to reduce person-to-person
   transmission of COVID-19 have been implemented to control the current outbreak and
   this has led to a significant reduction in the viral spread and a reduction in new
   cases. In Bulgaria, the disease began with a limited number of cases.
   The government took instant, timely and decisive measures, which led to a reduction in the
   spread of the infection. There was a weak first wave with a peak around April 20,
   when the new cases reached 91 per a day. There was a clear decrease in morbidity over
   the next month and a half, of which, due to the implemented measures. After restrictions were lifted and
   borders were opened, cases began to rise rapidly and right now Bulgaria is in ''second'' wave, with the intensity of morbidity of around of 250 new cases per a day.
  A very similar picture could be seen in the rest of Europe, where the mitigation of measures and lifting of the travel restrictions had also led to a rapid increase in the number of positive cases and morbidity rates.
A major part of the special efforts to predict, prevent and reduce
the transmission of the disease, is to develop a
mathematic-computation model for forecasting the COVID - 19 spread
in Bulgaria  in Europe and in the world wide.

\subsection{Epidemic Dynamics Modeling
and Analysis }

The modern approach for studying and forecasting the dynamics of
pandemic processes is connected to the use of deterministic
mathematical models described by differential equations. They track
the quantitative changes over time of main population groups
representing different stages of the course of the disease or
pandemic (virus carriers, hospitalized, recovered, etc.). The
equations in the mathematical models describe the consecution and
velocity of transitions from one group to another. The exact
dependencies in the equations are controlled by coefficients that
take into account the real (measurable) characteristics of the
infectious disease and the human population in the respective region
and are determined by the empirical data. The most commonly used
models, including for predicting the current COVID-19 pandemic, are
based on the so-called SIR and SEIR models, which are essentially
systems of nonlinear ordinary differential equations. Specifically
for COVID-19, see, for example, the Chris Murray Model
\cite{ChrisM}, the Epidemic Calculator \cite{EC}, and the works
\cite{NFer,NF2} of Neil Ferguson, Imperial College, London. The same
model has been used also to analyze the situation in China
\cite{Fang}. For more detailed and complied  analysis see a newly
published monograph \cite{Liu}.

The mathematical and computer modeling of the  COVID-19 pandemic is
a really interesting and important topic in which mathematical
modeling and high - performance  computational calculations clearly
cooperate. This is a modern tool not only for forecasting, but also
for studying the mechanisms of disease spread, for assessing the
effect of various interventions or strategies for controlling the
pandemic.

In  SEIR model the host population  is represented by the following
compartments:
\begin{itemize}
    \item Susceptible individuals $S(t)$.  These are the people who may be infected and can become
virus carriers. Usually at  the beginning of the pandemic, as in the
case on  COVID-19,  the whole population of the country is
susceptible.
    \item Exposed  individuals $E(t)$.
These are virus carriers  individuals in the latent stage, during
which they are not virus spreaders. They usually have no symptoms.
    \item Infectious individuals $I(t)$. These  are the
individuals who are with strong infectivity virus carriers and virus
spreaders. These are the people who may be passing the virus on to
others in case of contact. It is supposed that at the start there is
at least one infectious individual.
    \item Recovered individuals $R(t)$. These are the recovered  from
    the disease individuals (in the same group we calculate also the
    deceased individuals).
       \end{itemize}
The total population size is supposed to be a constant
$N=S(t)+E(t)+I(t)+R(t)$ during the considered  pandemic period.

In SEIR model  the quantitative changes over time are described by
the following Cauchy problem for system of nonlinear ordinary
differential equations (under assumptions that
 $S(t),\, E(t),\, I(t)$ and $R(t)$ are differentiable functions):
\begin{equation}\label{eq:SEIR}
\text{SEIR model:} \left|
\begin{array}{l}
\displaystyle{\frac{d S}{d t}}=- \frac{\beta}{N}S(t)I(t),\\\\
\displaystyle{\frac{d E}{d t}}=\frac{\beta}{N}S(t)I(t)-\omega E(t),\\\\
\displaystyle{\frac{d I}{d t}}=\omega E(t)-\gamma I(t),\\\\
\displaystyle{\frac{d R}{d t}}=\gamma I(t),\\\\
S(0)=S_0,\,E(0)=E_0,\,I(0)=I_0,\,R(0)=R_0,
\end{array}
\right.
\end{equation}
Here $S_0>0,\, E_0 \ge 0,\, I_0>0,\, R_0\ge 0$ are the initial
numbers of individuals in the fourth  compartments. The model
involves three  parameters:
    \begin{itemize}
        \item   Infection (transmission) rate $\beta$ of the biological epidemic,  which is a product of the
        of contact rate $r$ between people
in the population  and the probability of transmission $\beta_0$
upon contact between an Infectious individual and a Susceptible
individual;
\item Latent (incubation) rate
$\omega=\frac{1}{T_e},$  where $T_e$ denoted average latency time
(incubation period).
        \item   Rate of the Infectious  people transform to the Recovered  people, i.e.  recovery rate  $\gamma =\frac{1}{T_i},$ where
        $T_i$  denoted  the
        average  recovery time.
            \end{itemize}
A fundamental concept in epidemiology is the basic reproduction
number $\Re_0$, which measures the speed of spread of an infectious
disease. In SEIR model \eqref{eq:SEIR} (see
\cite{Driessche,Carcione}) it is
\begin{equation}\label{eq:R0seir}
\Re_0=\frac{\beta}{\gamma}.
\end{equation}
 When $\Re_0>1$ the speed of spread is positive
and SEIR model gives exponential growth of infectious individuals.
If
 $\Re_0=1$, then the speed of spread is zero (disease threshold).
If $\Re_0<1$ the spread speed is negative and  the infection
subsides.

             SEIR model \eqref{eq:SEIR} can be considered as generalization of the SIR model, suggested by
              Kermack and McKendrick \cite{Kermack}. SIR model has been widely applied in dynamic transmission
modeling of a directly transmitted pathogen such as influenza. The
latent period is ignored and the compartment of Exposed individuals
absents in it:
\begin{equation}\label{eq:SIR}
\text{SIR model:} \left|
\begin{array}{l}
\displaystyle{\frac{d S}{d t}}=- \frac{\beta_1}{N}S(t)I(t),\\\\
\displaystyle{\frac{d I}{d t}}=\frac{\beta_1}{N}S(t)I(t)-\gamma_1 I(t),\\\\
\displaystyle{\frac{d R}{d t}}=\gamma_1 I(t),\\\\
S(0)=S_0,\,I(0)=I_0,\,R(0)=R_0.
\end{array}
\right.
\end{equation}
Here the parameters are the infection rate $\beta_1$ and the
recovery  rate $\gamma_1$. The  basic reproduction number $\Re_0$ in
SIR model \eqref{eq:SIR} is defined as the ratio  of the infection
rate $\beta_1$ to recovery rate $\gamma_1.$

\begin{remark}\label{rem:NP}  Let us mention here the difference between both Models:
In SEIR model there are four different groups of people, instead of
three in SIR. This difference is very important because to solve the
inverse problem of finding the coefficients $\beta,\, \omega$ and
$\gamma$ in SEIR model we need more information, than in SIR, which
unfortunately is not public. With this reason to solve this problem
we are going in a different way (see Section 2). Let us also mention
some other models, which are more advanced. For example see SEIQRS
Epidemic Diffusion Model (see Liu, Cao, Liang, Chen monograph
\cite{Liu}, 2020). What is the difference with SEIR model for
example there? In SEIQRS model there is one more compartment of
Quarantined individuals and it is supposed that recovered people
have not immunity. If we look to the first equation in SIR or in
SEIR model, there is a product $S(t)I(t)$ which actually means that
the number of susceptible people depends on the number of meetings
between the all people $S(t)$ and all infectious cases $I(t)$.
Instead of this, in SEIQRS model \cite {Liu} this product has been
changed with the product $S(t)I_1(t)$, where $I_1(t)$ is the number
of another group of people, infective with strong infectivity, which
have not yet been quarantined. This model is obviously more
accurate; however, due to lack of publicly accessible data, this is
not applicable in Bulgaria for research purposes. If the researchers
had more accurate data on the number of quarantined individuals, it
would have been possible to make more accurate forecasts.
\end{remark}

\begin{figure}
\begin{center}
\begin{tabular}{ccc}
  \includegraphics[width=0.45\textwidth]{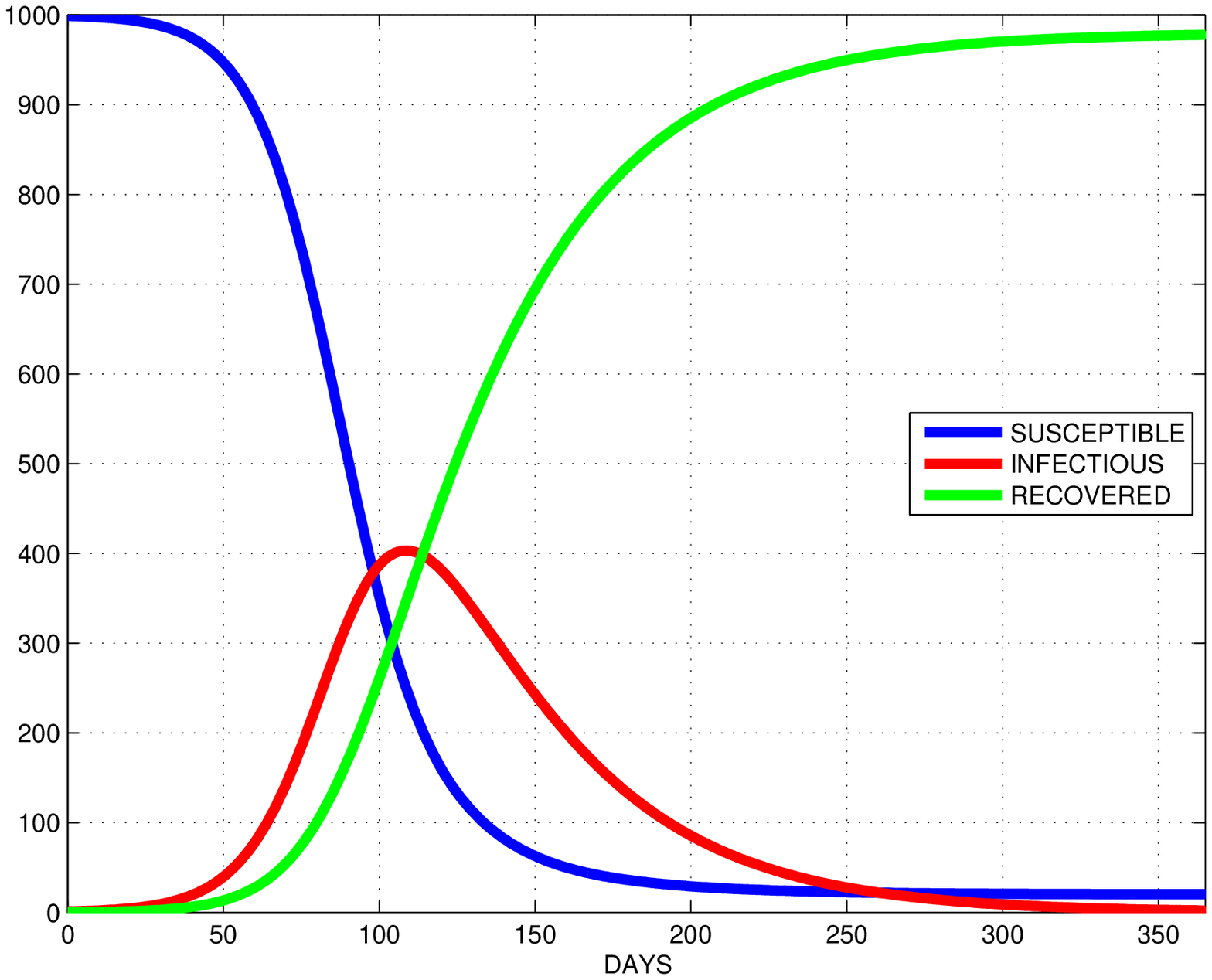}& &
  \includegraphics[width=0.45\textwidth]{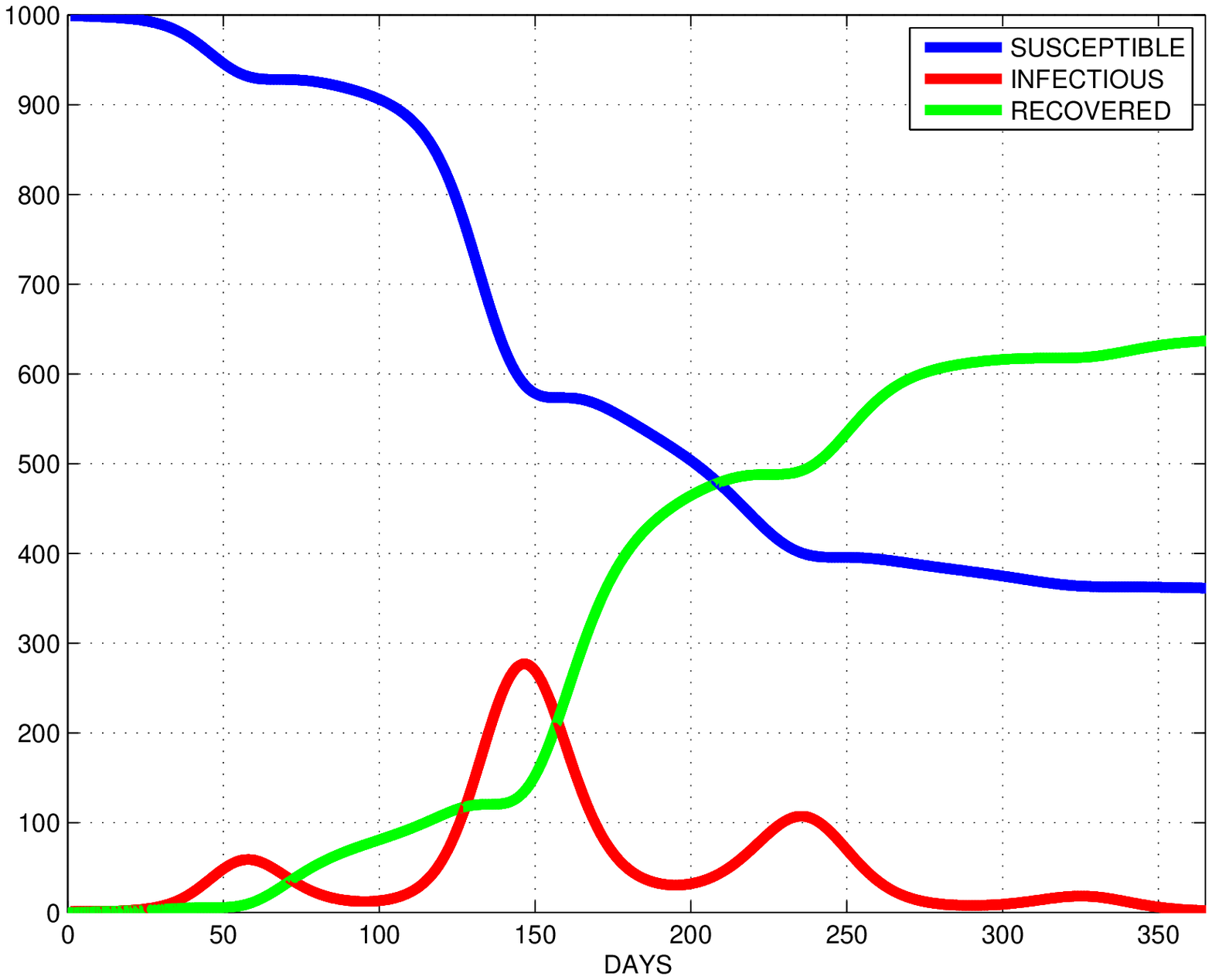} \\
  (a) constant coefficients & &(b) time-depended coefficients \\
\end{tabular}
            \end{center}
        \caption{SIR model infectious curve (in red) with one (a) or several (b) maxima.}
\label{fig:A}
    \end{figure}

The SIR and SEIR models with constant  coefficients are useful for
modeling of rapid spread of disease to a large number of people in a
given population within a short period of time (see \cite{White}).
Seasonal epidemics or controlled pandemic  by  a switching
prevention strategy, which may include vaccination and outbreak
control measures are described by models with time-depended
transmission parameters (see \cite{Ponciano,Chlanda,LiuSt} ). In
these cases the number of infectious individuals $I(t)$ can have
many local maxima (see Figure \ref{fig:A} (b)), instead of one pick
(in the cases of constant coefficients, Figure \ref{fig:A} (a)),
 and they can be controlled by good prevention strategy.
    As is mentioned in \cite{Fang} the value of the infection rate
$\beta$ can be changed by various government control measures. For
example, educational facilities closing,  gathering restrictions,
travel restrictions, cancelation of mass gatherings,  and similar
control measures impact on the values of of the rate $r$ of contact
between people in the population. On the other hand, the recovery
rate $\gamma$ can be changed, for example, by methodological
improvement on the diagnosis and treatment strategy. Therefore it is
reasonable to use time-depended models to describe the COVID-19
pandemic dynamics.

\subsection{Spatial Dynamics}

The spatial disease dynamics is among the challenging topics to
better understanding of future development of COVID-19 and other
related pandemics, due to the ever stronger impact of human mobility
to infectious disease dynamics on larger geographical scales. For
SEIR model, this means generalization of the basic system of
ordinary differential equations to systems of either integral
equations, partial differential equations or coupled ordinary
differential equations on a weighted graph, thus including the
spatial diffusion (see, e.g., \cite{Brokman-09, Jiang-19, Liu-19,
Takacs-20}). At a certain level of epidemic development, the long
range human mobility is responsible for the rapid geographical
spread of emergent infectious diseases, thus violating the Brownian
motion hypotheses. Although significantly more complex, the emerging
fractional diffusion models provide new opportunities for better
analysis of such non-local phenomena. The successful development of
spatial time-depended inverse SEIR models will require substantially
bigger amount and more reliable data to determine the related
coefficients, which are by default strongly heterogeneous in space.
The data for time-dependent density of individuals is just one of
the challenges. For example, in the numerical tests presented in
\cite{Brokman-09}, data for the flux of dollar bills in US is used
to parametrize the model of spatial dynamics. The analysis of
spatial dynamics of epidemic diseases is a topic with great social
and economic impact.  The common understanding of this is rapidly
increasing during the ongoing COVID-19 pandemic. This is
convincingly supported by the following example.  The Institute for
Health Metrics and Evaluation (IHME) is an independent research
center at the University of Washington, financially supported by the
Bill \& Melinda Gates Foundation \cite{IHME-20}.  For the next
decade, IHME has formulated the following ambitious target: ''Scale
up our geospatial analyses toward our aspiration of mapping all
diseases, risks, and covariates at 5 km by 5 km resolution or
better.'' In the spirit spatial SEIR models, such (or even coarser)
resolution can be achieved only via deep synergy of resources in
High Performance Computing (HPC) and Artificial Intelligence (AI) in
Big Data environment.

 The paper is organized as follows. In Section 2 we
introduce a time-depended inverse SEIR model for daily adjustment of
the infection, latent, and the recovery rates. Additionally, we
propose difference schemes for numerically solving the corresponding
direct and inverse problems. In Section 3 this theoretical machinery
is used for experimental analysis of the pandemic situation in
Bulgaria and for a development of a strategy for mid-time-frame
forecasting. Numerical results for three different 14-days-periods
are documented and analyzed. Conclusions are outlined in Section 4.

\section{COVID-19 spread modeling  by time-depended inverse SEIR model}

In this section we consider SEIR model with time depended
parameters:
\begin{equation}\label{eq:TDSEIR}
\text{TD-SEIR model:} \left|
\begin{array}{l}
\displaystyle{\frac{d S}{d t}}=- \frac{\beta(t)}{N}S(t)I(t),\\\\
\displaystyle{\frac{d E}{d t}}=\frac{\beta(t)}{N}S(t)I(t)-\omega(t) E(t),\\\\
\displaystyle{\frac{d I}{d t}}=\omega(t) E(t)-\gamma(t) I(t),\\\\
\displaystyle{\frac{d R}{d t}}=\gamma(t) I(t),\\\\
S(0)=S_0,\,E(0)=E_0,\,I(0)=I_0,\,R(0)=R_0.
\end{array}
\right.
\end{equation}
The parameters in TD-SEIR model \eqref{eq:TDSEIR} are
 different in different stages of pandemic  and they
are specifically for each country. We assume that
$\omega(t)=1/T_e(t) \le 1$ and $\gamma(t)=1/T_i(t)\le 1,$ where
$T_e(t)$ and $T_i(t)$ are average latency time and average recovery
time to the moment $t$.

One can solve the Cauchy problem in  TD-SEIR model \eqref{eq:TDSEIR}
knowing  the initial data and the value of the parameters. In such
way functions describing the quantitative changes in considered
compartment of population can be found.  But this is not the case.
Official data sources  relating to the COVID-19 pandemic
  contains information only for number of active cases
 and number of Recovered individuals
$R(t)$. Since many people who have the disease show no symptoms it
is reasonable to assume that active
 cases  are infectious and exposed individuals and will denote
 them by $A(t)=E(t)+I(t).$ The parameters in the TD-SEIR model \eqref{eq:TDSEIR} are
unknown and therefore in order to study the COVID-19 pandemic
dynamics the first step is to solve

\smallskip

\textbf{The inverse problem: }To find the parameters $\beta(t),\,
\omega(t)$ and $\gamma(t)$ in the TD-SEIR  model \eqref{eq:TDSEIR}
using the knowing real data for the functions $A(t)=E(t)+I(t)$ and
$R(t)$.

\smallskip

 Actually the confirmed data for
$A(t)$ and $R(t)$ are not enough for
  analytically solvability of the inverse problem. As we have mentioned in
  Remark \ref{rem:NP},  to solve the inverse problem in SEIR model we need more information,
  than in SIR model, which unfortunately is not public.
  With this reason  we fix the latency rate $\omega(t)$ to be a constant during
the pandemic period. The sensitivity analysis conducted in
\cite{Fang} in case of knowing real data for $I(t)$ (instead of
$A(t)$) shows that the change of $\gamma(t)$ had strong impact on
the infected population in China in contrast to $\omega(t)$.
 Then  we are able to find $\beta(t)$ and $\gamma(t)$
in TD-SEIR model \eqref{eq:TDSEIR}.

We will  show how  the parameters in  TD-SEIR model
\eqref{eq:TDSEIR} can be calculated using known real data for
COVID-19 disease and solving the corresponding inverse problems.
Different procedure for calculation of the parameters in SIR/SEIR
models and different data fitting procedures like interpolation
splines modeling or least square estimation are used in
\cite{Chen,Kounchev,Ma}. We propose different data fitting method.

\begin{remark}Let us mention that under the term ''real data'' we use the official data,
 given by many public sources. Obviously, we use them instead of others which
 no one knows exactly!
\end{remark}

In order to solve the formulated inverse  problem  we introduced the
new functions
\begin{equation}\label{eq:Ibg2}
I_{\gamma}(t)=\gamma(t)I(t),\; I_{\beta}(t)= \beta(t)I(t)
\end{equation}
and summed up the second and  the third equations in TD-SEIR model
\eqref{eq:TDSEIR}:
\begin{equation}\label{eq:SAIR}
 \left|
\begin{array}{l}
\displaystyle{\frac{d S}{d t}}=- \frac{1}{N}I_{\beta}(t)S(t),\\\\
\displaystyle{\frac{d A}{d t}}=\frac{1}{N}I_{\beta}(t)S(t)-I_{\gamma}(t),\\\\
\displaystyle{\frac{d I}{d t}}=\omega(t)(A(t)-I(t))- I_{\gamma}(t),\\\\
\displaystyle{\frac{d R}{d t}}=I_{\gamma}(t),\\\\
S(0)=S_0,\,A(0)=A_0,\,I(0)=I_0,\,R(0)=R_0.
\end{array}
\right.
\end{equation}

Now we are able to solve the inverse problem by the following steps:

\begin{enumerate}

\item  We consider period of $n \in \mathbb{N}$ days of the pandemic.
Let us denote by $t_1,\,t_2, \ldots , t_n$  the days of this
 period.

\item The available dataset contains data for the confirmed  Active cases $A_k=A(t_k)$
 and Recovered
individuals $R_k=R(t_k)$ at the day $t_k,\, k =1,2,\ldots,n$.
Therefore the number of Susceptible individuals is also known
$S_k=S(t_k)=N-A_k-R_k.$ But the numbers of the Exposed individuals
$E_k=E(t_k)$ and Infectious individuals $I_k=I(t_k)$ are unknown
(actually, only the sum $E_k+I_k=A_k$ is known).

\item  Let us denote by $\beta_k=\beta(t_k), \omega_k=\omega(t_k)$ and $ \gamma_k=\gamma(t_k)$  the
unknown  values of the  parameters in TD-SEIR model at the day
$t_k,\, k =1,2,\ldots,n-1.$  It is naturally to assume  that
$\beta_k,\, \omega_k$ and $\gamma_k$ are nonnegative constants.

\item We  rewrite  TD-SEIR model \eqref{eq:TDSEIR}  as the following difference
problem:
\begin{equation}\label{eq:DTDSEIR}
\mathbf{DTD-SEIR_k:} \left|
\begin{array}{l}
S_k-S_{k-1}=- \frac{1}{N}I_{\beta,k-1}S_{k-1},\\\\
A_k-A_{k-1}=\frac{1}{N}I_{\beta,k-1}S_{k-1}-I_{\gamma,k-1},\\\\
I_{k}-I_{k-1}=\omega_{k-1}(A_{k-1}-I_{k-1})- I_{\gamma,k-1},\\\\
R_k-R_{k-1}=I_{\gamma,k-1},
\end{array}
\right. ,\,k=2,3,\ldots n.
\end{equation}
where
\begin{equation}\label{eq:Ibg}
\begin{aligned}
I_{\beta,k}&=I_{\beta}(t_k)=\beta_kI_k,\\
I_{\gamma,k}&=I_{\gamma}(t_k)=\gamma_kI_k.
\end{aligned}
\end{equation}
It is natural to assume that  at the start of the pandemic all
Active cases are Infectious individuals $A_1=I_1$.

\item From the first and the last equations in problem $DTD-SEIR_k$ \eqref{eq:DTDSEIR} we have
\begin{equation}\label{eq:Ibgk}
I_{\beta,k-1}=-N\frac{S_k-S_{k-1}}{S_{k-1}},\;I_{\gamma,k-1}=R_k-R_{k-1},\,
k=2,3,\ldots n.
\end{equation}

\item From the third equation in problem $TDT-SEIR_k$ \eqref{eq:DTDSEIR} we obtain
 \begin{equation}\label{eq:Ik}
 I_k=(1-\omega_{k-1})I_{k-1}+\omega_{k-1}A_{k-1}-I_{\gamma,k-1} \ge 0,\, k=2,3,\ldots n,
 \end{equation}
from which follows \begin{equation}\label{eq:omega}
(A_{k-1}-I_{k-1})\omega_{k-1} \ge I_{\gamma,k-1}-I_{k-1},\,
k=2,3,\ldots n.
\end{equation}
\item Since  $\gamma_{k-1} \le 1$ we have $I_{\gamma,k-1} \le I_{k-1}$ and taking into account
the inequality $A_{k-1}\ge I_{k-1}$ we obtain that condition
\eqref{eq:omega} is fulfilled. This allow us to set
$\omega_{k-1}=1/T_e$, where $T_e=const.$ is the average incubation
period for COVID-19. Now we can calculate the values
\begin{equation}\label{eq:bgk}
\beta_k=\frac{I_{\beta,k}}{I_k},\,
\gamma_k=\frac{I_{\gamma,k}}{I_k},\, k=1,2,\ldots n-1.
\end{equation}
\end{enumerate}

 \begin{remark}The obtained values of  $I_{\beta,k}$ and $I_{\gamma,k}$ in \eqref{eq:Ibgk} allow us
to calculate the value of the basic reproduction number at the day
$t_k$ if $I_{\gamma,k}>0:$
\begin{equation}\label{eq:R0}
\Re_{0,k}=\frac{\beta_k}{\gamma_k}=\frac{I_{\beta,k}}{I_{\gamma,k}},\,
k=1,2,\ldots n-1.
\end{equation}
\end{remark}

 Using  TD-SEIR model \eqref{eq:SAIR} and the obtained values for parameters $\beta =(\beta_1,\beta_2\ldots,
\beta_{n-1}),$ $\omega =(\omega_1,\omega_2\ldots, \omega_{n-1})$ and
$\gamma =( \gamma_1,\gamma_2,\ldots, \gamma_{n-1})$ we  solve
numerically recurrent sequence of initial problems  for differential
equation systems that corresponded to the  days $t_1,t_2\ldots t_n$
of the considered period of the pandemic. More precisely for $t\in
[t_{k-1},t_{k}],\;k=2,3,\ldots,n,$  we solve the Cauchy problem
\begin{equation}\label{eq:Cauchyk}
\mathbf{TD-SEIR_k:} \left|
\begin{array}{l}
\displaystyle{\frac{d W_k}{d t}}=- \frac{\beta_{k-1}}{N}W_{k}(t)Y_{k}(t),\\\\
\displaystyle{\frac{d X_k}{d t}}= \frac{\beta_{k-1}}{N}W_{k}(t)Y_{k}(t) - \omega_{k-1} X_{k}(t),\\\\
\displaystyle{\frac{d Y_{k}}{d t}}=\omega_{k-1} X_{k}(t)-\gamma_{k-1} Y_{k}(t),\\\\
\displaystyle{\frac{d Z_{k}}{d t}}=\gamma_{k-1} Y_{k}(t),\\\\
W_k(t_{k-1})=W_{k-1}(t_{k-1}),\,X_k(t_{k-1})=X_{k-1}(t_{k-1}),\,Y_k(t_{k-1})=Y_{k-1}(t_{k-1}),\,Z_k(t_{k-1})=Z_{k-1}(t_{k-1}),
\end{array}
\right.
\end{equation}
where $W_1(t_1)=S_1,\,X_1(t_1)=E_1=0,\,Y_1(t_1)=I_1=A_1$ and
$Z_1(t_1)=R_1.$
 The value of the
functions $W_{k-1}(t),$ $X_{k-1}(t),$ $Y_{k-1}(t),$ $Z_{k-1}(t)$ (
solution of the problem $TD-SEIR_{k-1}$) at the end of the day
$t_{k-1}$ are used as initial data in the next problem $TD-SEIR_k,$
that describes the quantitative changes over the day $t_k.$

 In the next section we apply the method described above to the Bulgarian
  COVID-19 data and
  approximate values of Active cases, New cases per day
 and Recovered individuals.

\section{Application of the TD- SEIR model to the Bulgarian COVID-19 data}

\subsection{Data}

The data of COVID-19 in this study were mainly obtained from
\begin{itemize}
\item
Johns Hopkins University:

https://coronavirus.jhu.edu/;
 \item
HDX Humanitarian Data Exchange:

https://data.humdata.org/dataset/novel-coronavirus-2019-ncov-cases;
 \item
 The official Bulgarian Unified  Information Portal:

 https://coronavirus.bg/
\end{itemize}

\noindent  Population: $N = 7 000 000.$

\noindent The first reported case: on March 8, 2020.

\noindent The period under consideration: $t_1=$ March 8, 2020, ...,
$t_n=$ July 12, 2020.

\subsection{Methods and Results}

\begin{figure}[b]
\begin{center}
\begin{tabular}{ccc}
  \includegraphics[width=0.45\textwidth]{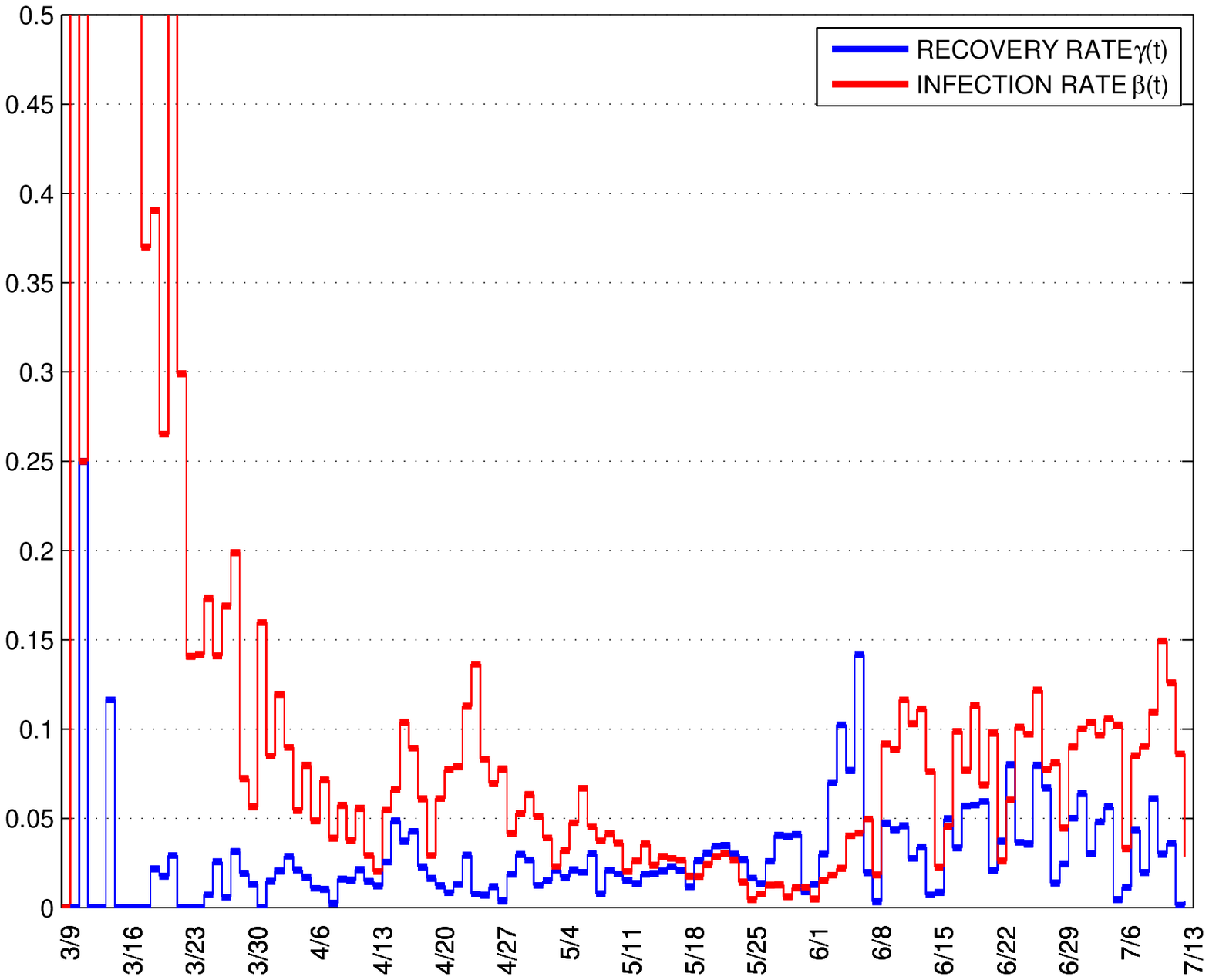}&
  & \includegraphics[width=0.45\textwidth]{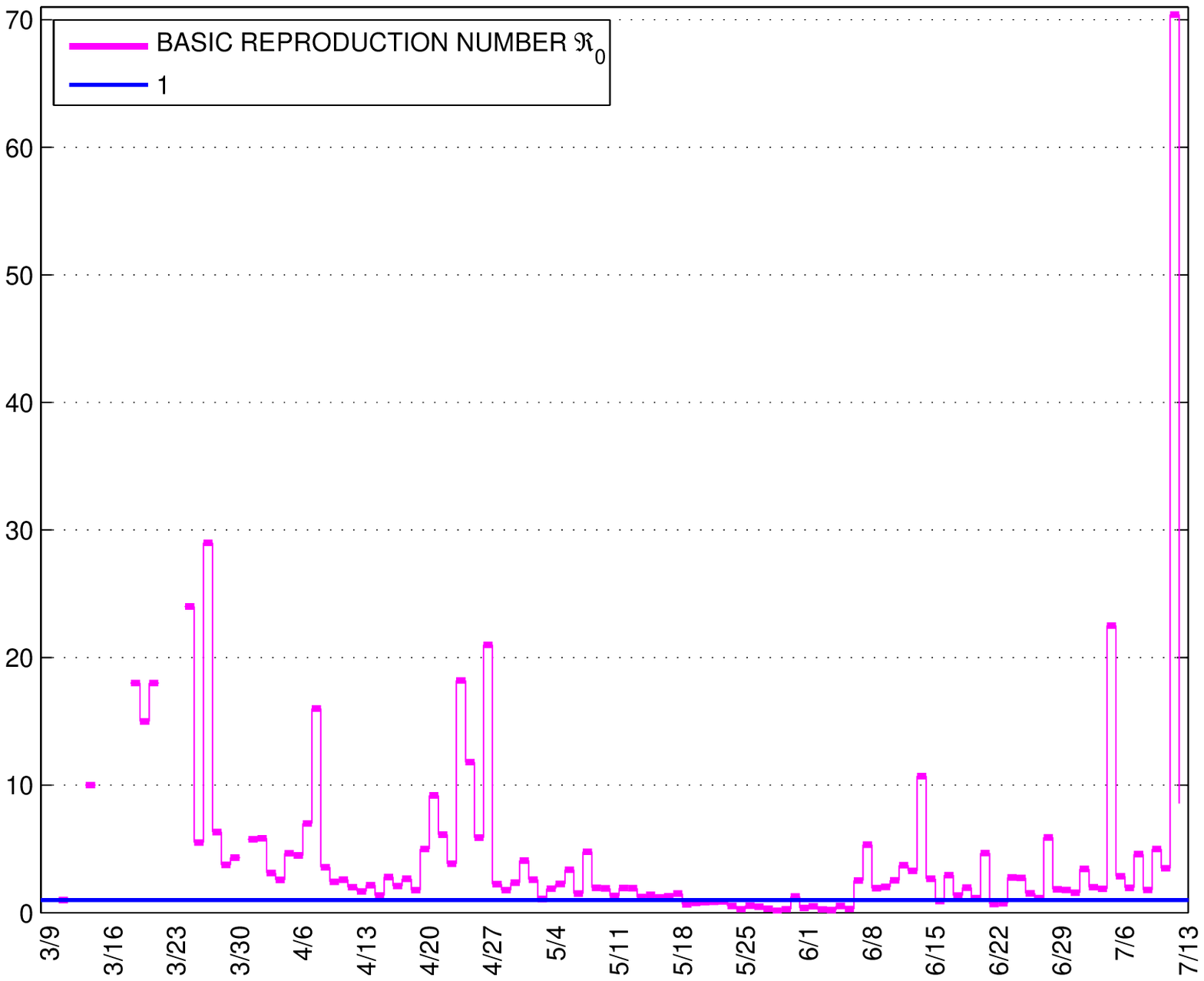}\\
  (a) the parameters $\beta(t)$  and $\gamma(t)$ & & (b) the basic reproduction number $\Re_0(t)$
\end{tabular}
            \end{center}
        \caption{Calculated values of SEIR model's parameters and basic reproduction number for Bulgaria.}
        \label{fig:BGSEIRparam}
    \end{figure}

\begin{figure}[]
\begin{center}
            \begin{tabular}{c}
\includegraphics[width=0.7\textwidth]{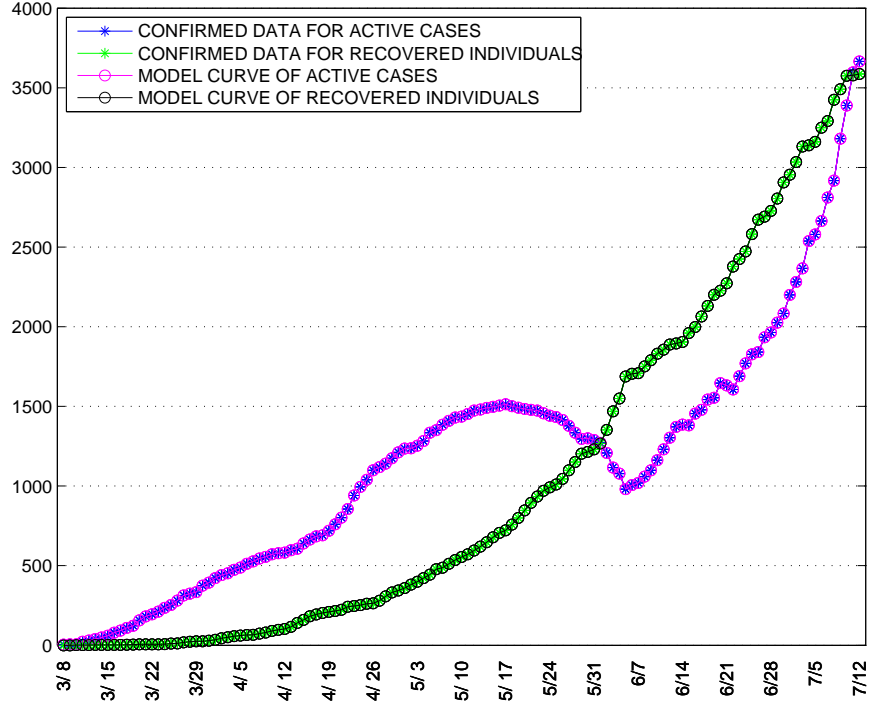}
        \end{tabular}
        \end{center}
        \caption{ Confirmed data and SEIR model's curves for Bulgaria.}
        \label{fig:BGSEIR}
    \end{figure}

To calculate the values of the infection rate $\beta(t)$ and the
recovery rate $\gamma(t)$ during the considered period  with
DTD-SEIR model \eqref{eq:DTDSEIR} we take average incubation period
in Europe $T_e=4$ (see \cite{Bohmer}) and set $\omega_k=1/T_e$.
Additionally, we assume that the data, provided by the official
 Bulgarian Unified Information Portal corresponds to all active cases
 (the union of exposed and infectious individuals) with a small time shift
 of $T_e$ days. Indeed, most of the time the pandemic in Bulgaria had a
 clustered spreading profile and most of the daily tested individuals
 were first-hand contacts of already confirmed cases. Therefore,
 it is hard to judge when exactly they were infected, e.g., were they
 only exposed or already infectious $T_e$ days ago. The situation with
 the recovered individuals is similar (one is considered ''recovered'' after
 a negative test, but the testing itself might have happen several days after
 the actual recovery). Only for the dead individuals the exact date is known,
 but their daily percentage is significantly low and can be neglected. Thus, we
 decided not to introduce time shifts in the numerical experiments and accept the
 convention that the official data resembles the active cases and the recovered
 individuals for the day they were announced. The
obtained values $\beta_k$ and $\gamma_k$ corresponding to the days
of the considered period are shown on the Figure
\ref{fig:BGSEIRparam} (a). The calculated values of the basic
reproduction number $\Re_{0,k}$ (when $\gamma_k>0$) are shown on
Figure \ref{fig:BGSEIRparam} (b).
    To verify  TD-SEIR model \eqref{eq:SAIR} we solve recurrent sequence
of initial problems \eqref{eq:Cauchyk} with obtained values
$\beta_k$ and $\gamma_k$ ($\omega_k$ are fixed) for parameters. On
the Figure \ref{fig:BGSEIR} are shown the real data for active cases
$A_k$ and Recovered individuals $R_k$ during the considered period,
and the models curves, obtained using the calculated values
$X_k(t_k)+Y_k(t_k)$ (for active cases) and $Z_k(t_k)$ (for recovered
individuals). We observe very good agreement between the calculated
and the real data. All numerical simulations were performed using
Matlab developed by MathWorks \cite{Matlab}.
We perform three sets of numerical experiments, related to
predicting a 14-day-time-frame in the future, based on the available
official data for Bulgaria up to the first day of the time frame.
Those time frames are: April 20 -- May 03, June 29 -- July 12, and
July 13 -- July 26, respectively. The first time frame is related to
the peak of the first wave of the virus. The second time frame is
related to the last known date. The third time frame is a forecast
for the future, thus we cannot compare the numerical results with
official data.

To fix the notation, let the first day of the forecast period be
denoted by $t_K$. Then, the last day of the forecast period will be
$t_{K+13}$. We assume that the information, provided in the morning
of every day $t_{k+1}$ is related to parameter values of the day
$t_k$, $k=1,2,\dots,n-1$, e.g., the $79$ reported new cases on June
10 are considered as the $C_k$ value for June 9. We always solve the
Cauchy problem \eqref{eq:Cauchyk} with initial data $\{S_{K-1},
A_{K-1}, I_{K-1}, R_{K-1}\}$ and parameter $ \omega=0.25,
\beta_{K-2}, \gamma_{K-2}$. The prediction of the parameters
$\{\beta_k\}_{K-1}^{K+12}$ and $\{\gamma_k\}_{K-1}^{K+12}$ is
performed on two levels. The first level uses the six ratios
$\{\beta_k/\beta_{k+1}\}_{K-9}^{K-4}$, respectively
$\{\gamma_k/\gamma_{k+1}\}_{K-9}^{K-4}$ from the week, before the
time frame. The same ratios are used for both corresponding
parameters from the first and second week of the forecast, e.g.,
$$
\frac{\tb_{K+5}}{\tb_{K+6}}=\frac{\beta_{K-2}}{\tb_{K-1}}=\frac{\beta_{K-9}}{\beta_{K-8}},\qquad
\frac{\tg_{K+5}}{\tg_{K+6}}=\frac{\gamma_{K-2}}{\tg_{K-1}}=\frac{\gamma_{K-9}}{\gamma_{K-8}},
$$
and analogously for the other five ratios. Note, that we denote by
$\tb_k$ and $\tg_k$ the predicted values of the parameters, using
the proposed methodology, in order to distinguish them from the
corresponding official ground-true data $\beta_k$ and $\gamma_k$,
reported by the Bulgarian officials. We have tried several different
approaches for optimal fitting of the parameter ratios, namely
various convex combinations of the corresponding ratios for a longer
time frame in the past (up to four weeks, see documented data in
Table~\ref{tab:beta_ratio}), but the best results were obtained by
simple repetition of the ratios from the very last week, as
described above. Possible explanations are related to the rapid
changes in the government control measures and the continuous growth
of licensed laboratories for PCR testing, which gradually increases
the overall tested people every week.
\begin{table}
 \caption{\label{tab:beta_ratio}Documentation of the
ratio $\beta_{k-2}/\beta_{k-1}$ on a weekly basis.}
\begin{tabular}{|c|c|c|c|c|c|c|}\hline
week & Mon/Tue & Tue/Wed & Wed/Thur & Thur/Fri & Fri/Sat &
Sat/Sun\\\hline
3/9 -- 15& 2.000 & 0.054 & 2.357 & 1.667 & 1.506 & 1.367\\
3/16 -- 22& 1.913 & 0.948 & 1.473 & 0.506 & 1.753 & 2.124\\
3/30 -- 4/5& 0.820 & 1.227 & 0.835 & 0.849 & 2.752 & 1.278\\
4/6 -- 12& 1.884 & 0.710 & 1.331 & 1.648 & 0.683 & 1.641\\
4/13 -- 19& 1.839 & 0.677 & 1.524 & 0.677 & 1.904 & 1.446\\
4/20 -- 26& 0.832 & 0.636 & 1.163 & 1.463 & 2.087 & 0.478\\
4/27 -- 5/3& 0.980 & 0.700 & 0.826 & 1.639 & 1.196 & 0.896\\
5/4 -- 10& 0.787 & 0.837 & 1.238 & 1.311 & 1.686 & 0.724\\
5/11 -- 17& 0.714 & 1.479 & 1.204 & 0.910 & 1.136 & 1.798\\
5/18 -- 24& 0.732 & 1.502 & 0.826 & 1.036 & 1.034 & 1.511\\
5/25 -- 31& 0.726 & 0.840 & 0.941 & 1.128 & 1.888 & 3.166\\
6/1 -- 7& 0.591 & 0.990 & 2.071 & 0.557 & 0.972 & 2.343\\
6/8 -- 14& 0.847 & 0.829 & 0.543 & 0.964 & 0.842 & 2.723\\
6/15 -- 21& 1.033 & 0.763 & 1.131 & 0.924 & 1.459 & 3.357\\
6/22 -- 28& 0.459 & 1.285 & 0.679 & 1.645 & 0.705 & 3.734\\
6/29 -- 7/5& 0.596 & 1.042 & 0.798 & 1.572 & 0.956 & 1.814\\
7/6 -- 12& 0.900 & 0.964 & 1.073 & 0.913 & 1.037 & 3.089\\\hline
\end{tabular}
\end{table}
\begin{figure}[]
\begin{center}
\includegraphics[width=0.75\textwidth]{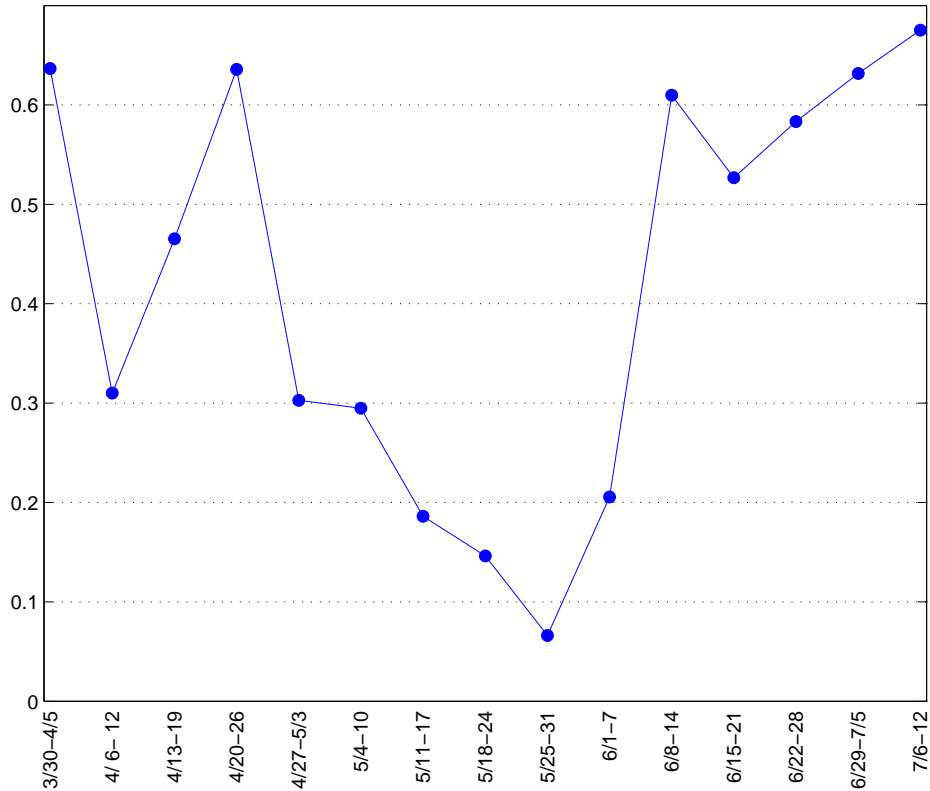}
\end{center}
\caption{The accumulated data $\beta_{week_i}$}
\label{fig:beta_week}
\end{figure}
The second level is to approximate the accumulative week parameters
$\{\beta_{week_i}, \beta_{week_{i+1}}\}$, respectively
$\{\gamma_{week_i}, \gamma_{week_{i+1}}\}$ for the weeks: $week_i$
-- the days between $t_K$ and $t_{K+6}$, and $week_{i+1}$ -- the
days between $t_{K+7}$ and $t_{K+13}$. For the first week, we have
the obvious relation:
\begin{equation}\label{eq:week prediction}
\tb_{week_i}:=\sum_{k=K-1}^{K+5}\tb_k,\qquad
\tg_{week_i}:=\sum_{k=K-1}^{K+5}\tg_k.
\end{equation}

For the second week, we fit $\beta_{week_{i+1}}$ and
$\gamma_{week_{i+1}}$, using the available week information from
several previous weeks. We used one fitting algorithm for the first
time-frame and another fitting algorithm for the remaining two
time-frames, due to particularities of the overall situation in
Bulgaria (see Fig.~\ref{fig:beta_week}). Further, once we predict
$\tb_{week_{i+1}}$ and $\tg_{week_{i+1}}$, together with the ratios
$\{\tb_k/\tb_{k+1}\}_{K+6}^{K+11}$
$\{\tb_k/\tb_{k+1}\}_{K+6}^{K+11}$, we can straightforwardly invert
\eqref{eq:week prediction} and derive $\{\tb_k\}_{K+6}^{K+12}$,
$\{\tg_k\}_{K+6}^{K+12}$. Finally, having all the necessary values
$\{\tb_k\}_{K-1}^{K+12}$ and $\{\tg_k\}_{K-1}^{K+12}$, we can solve
the Cauchy problem TD-SEIR$_k$ along the whole time period of
interest.

For the first 14-days-time-frame April 20 -- May 03, based on the
data in Figure \ref{fig:beta_week} we predicted $\tb_{week_{i+1}}$
and $\tg_{week_{i+1}}$  via
$$
\tb_{week_{i+1}}=\tb_{week_i}+\beta_{week_{i-2}}-\beta_{week_{i-3}},\qquad
\tg_{week_{i+1}}=\frac{\tg_{week_i}+\gamma_{week_{i-1}}+\gamma_{week_{i-2}}+\gamma_{week_{i-3}}}{4}.
$$
We used a parallelogram rule for the infection rate $\beta$
prediction, due to the strict government control measures at that
time (facilities and parks closing, social distancing and travel
restrictions, etc.). Thus, in general, the weekly infection rate was
expected to monotonically decay towards zero. However, this was not
the case for two consecutive weeks with linear growth of the
parameter (April 13 -- April 19 and April 20 -- April 26), which was
mainly due to the two big Orthodox holidays in Bulgaria: Palm Sunday
(April 12) and Easter (April 19). The strict measures quickly
inverted the monotonic behavior of $\beta$ and the weekly values for
April 27 -- May 3 were already in vicinity of those before the
holidays (from April 6 -- April 12). We used a simple averaging rule
for the recovery rate $\gamma$ prediction, due to the early stage of
the pandemic and the very long recovery period for infected patient.
Therefore, practically most of the recovered individuals, counted
within this period, were the deceased ones. This number behaved
quite stably and did not oscillate much, meaning that it seemed
natural to approximate it with a constant function.

\begin{figure}[t]
\begin{center}
\begin{tabular}{ccc}
Active cases & Daily cases & Recovered individuals\\
\includegraphics[width=0.32\textwidth]{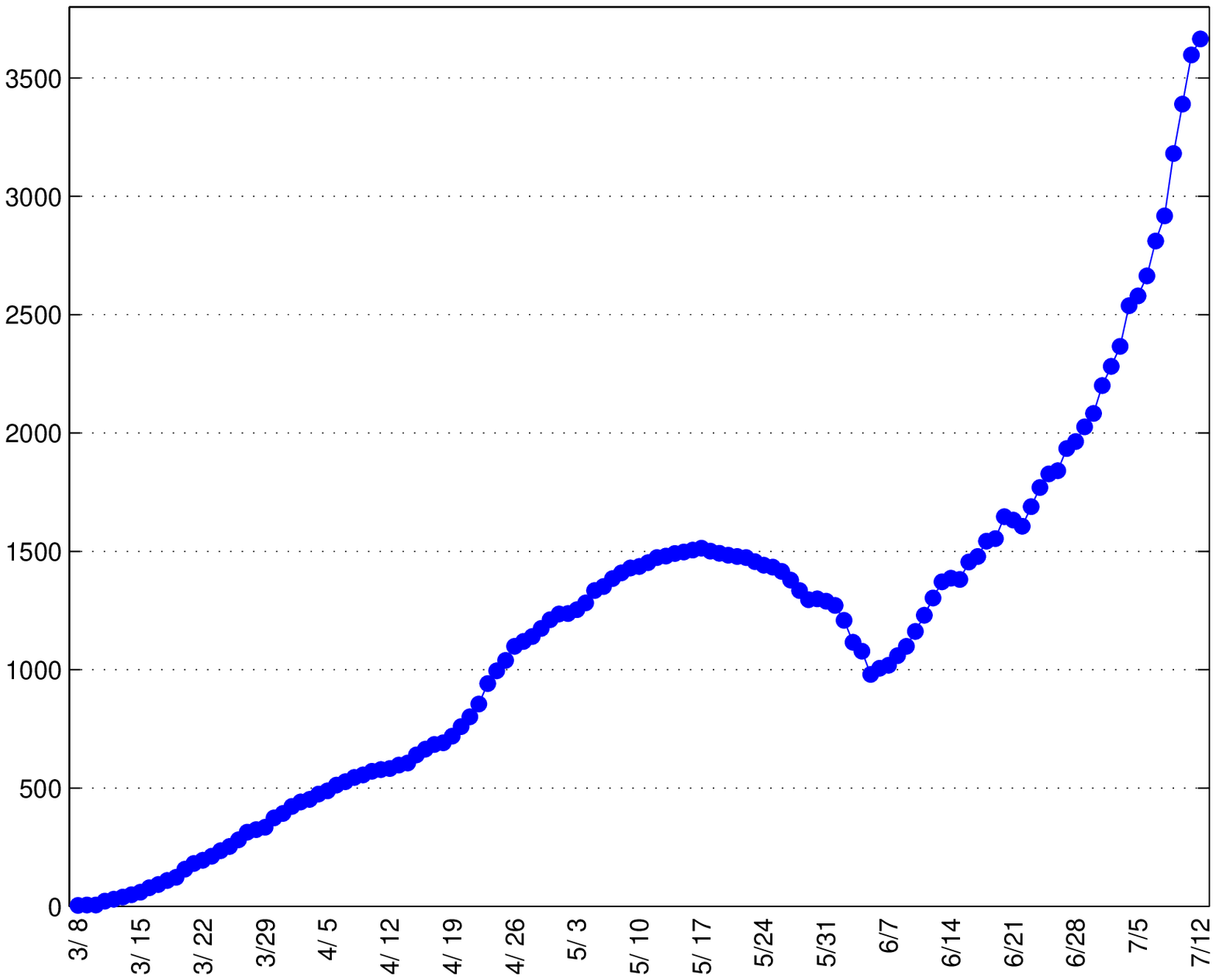} &
\includegraphics[width=0.32\textwidth]{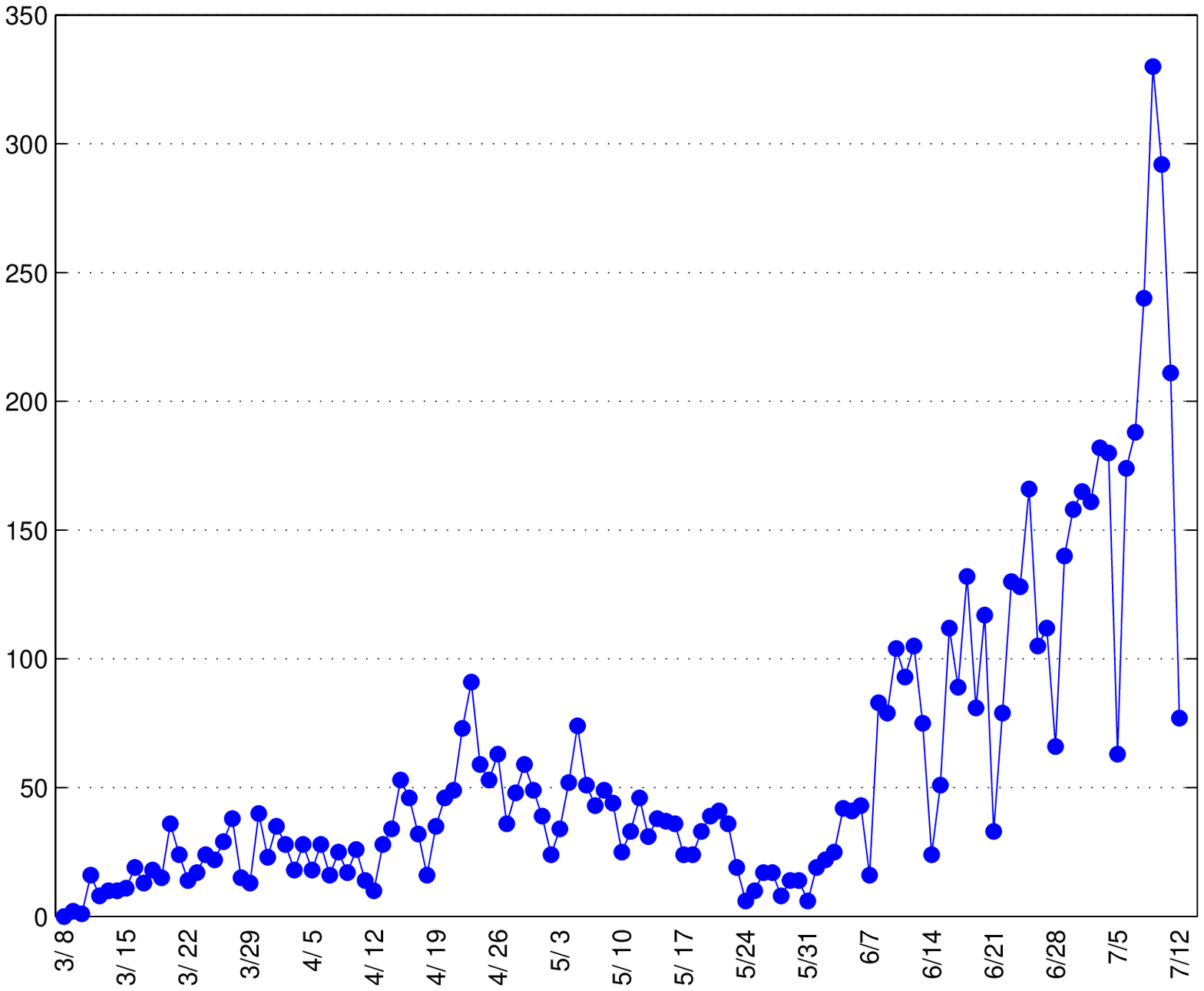} &
\includegraphics[width=0.32\textwidth]{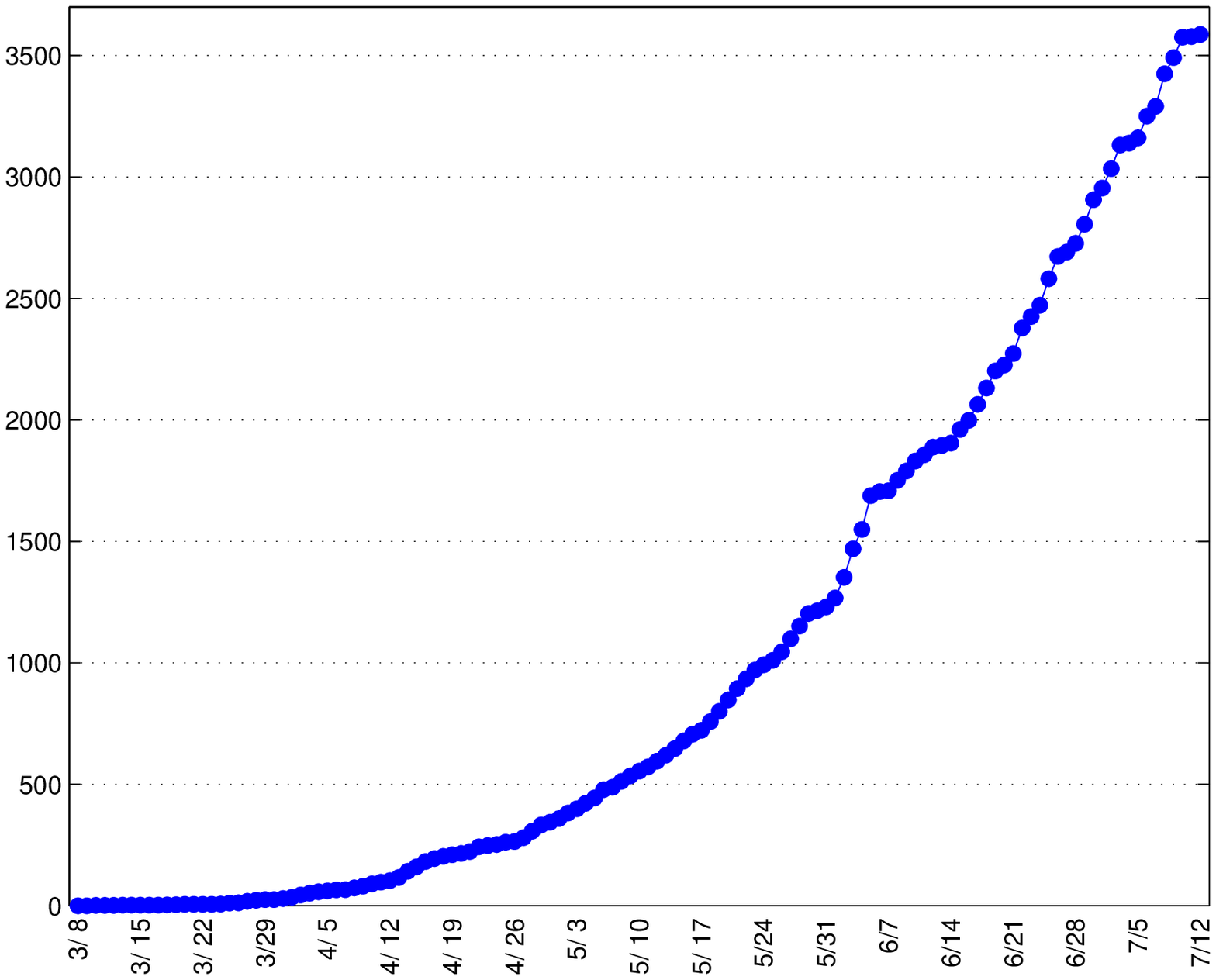} \\
\includegraphics[width=0.32\textwidth]{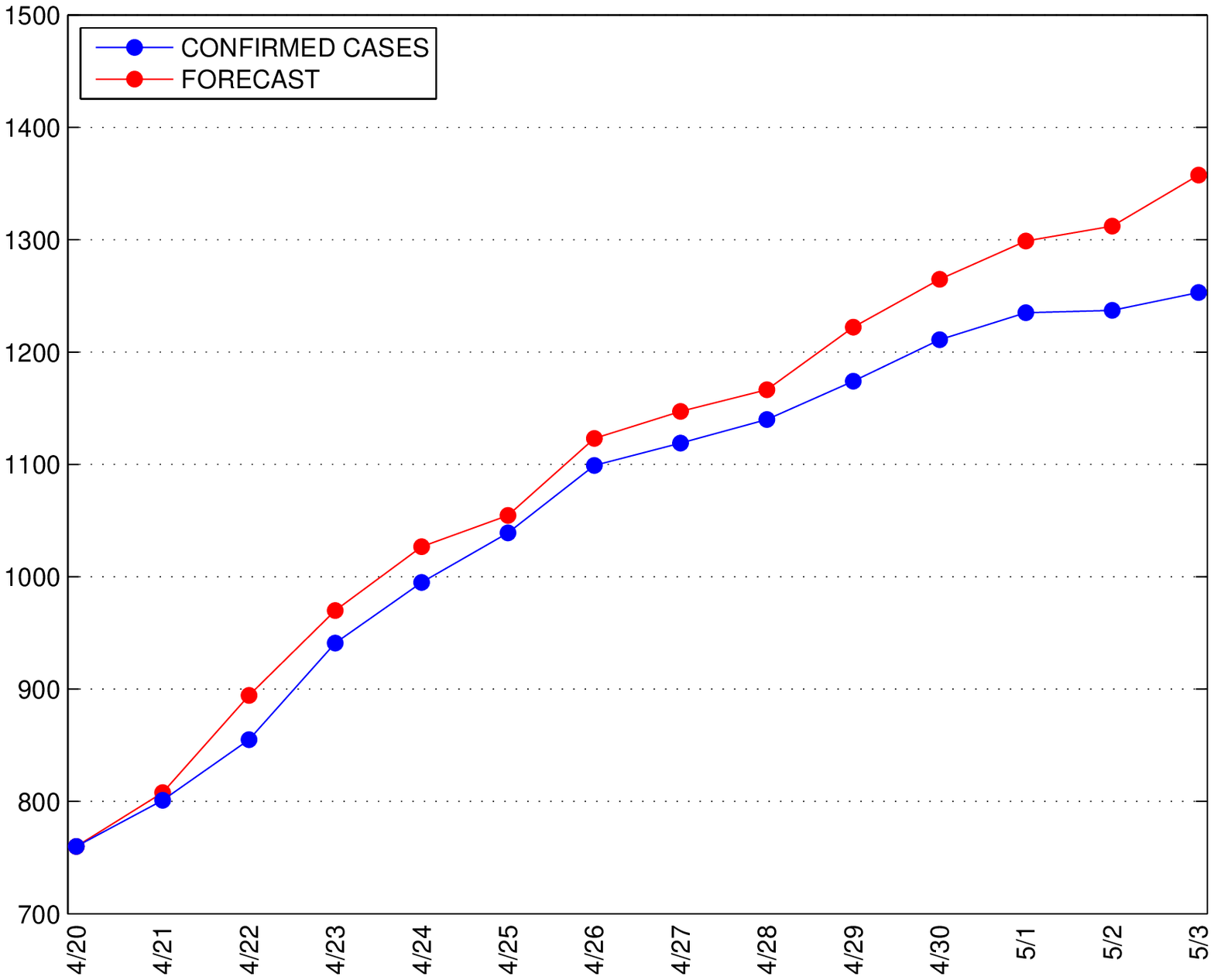} &
\includegraphics[width=0.32\textwidth]{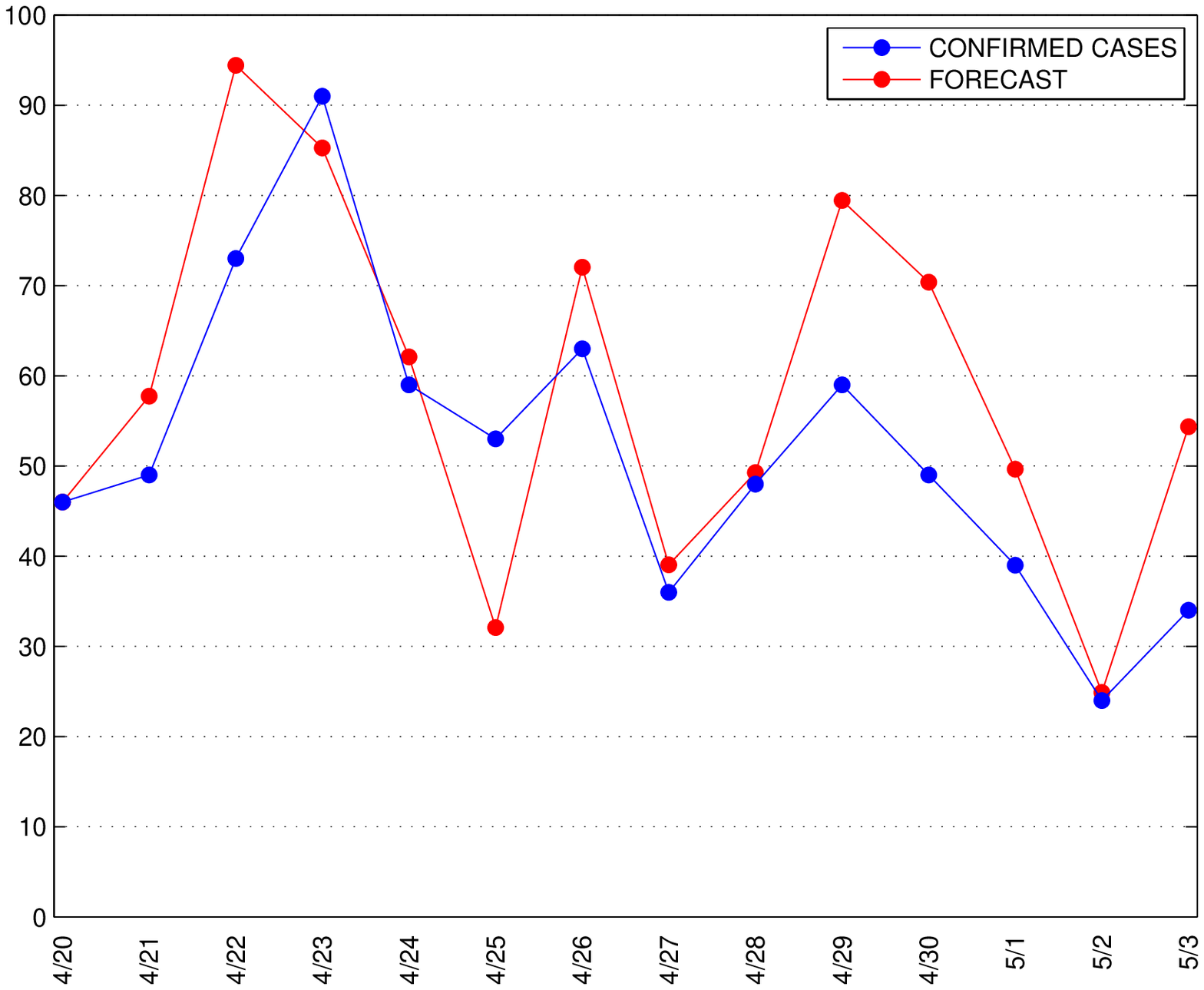} &
\includegraphics[width=0.32\textwidth]{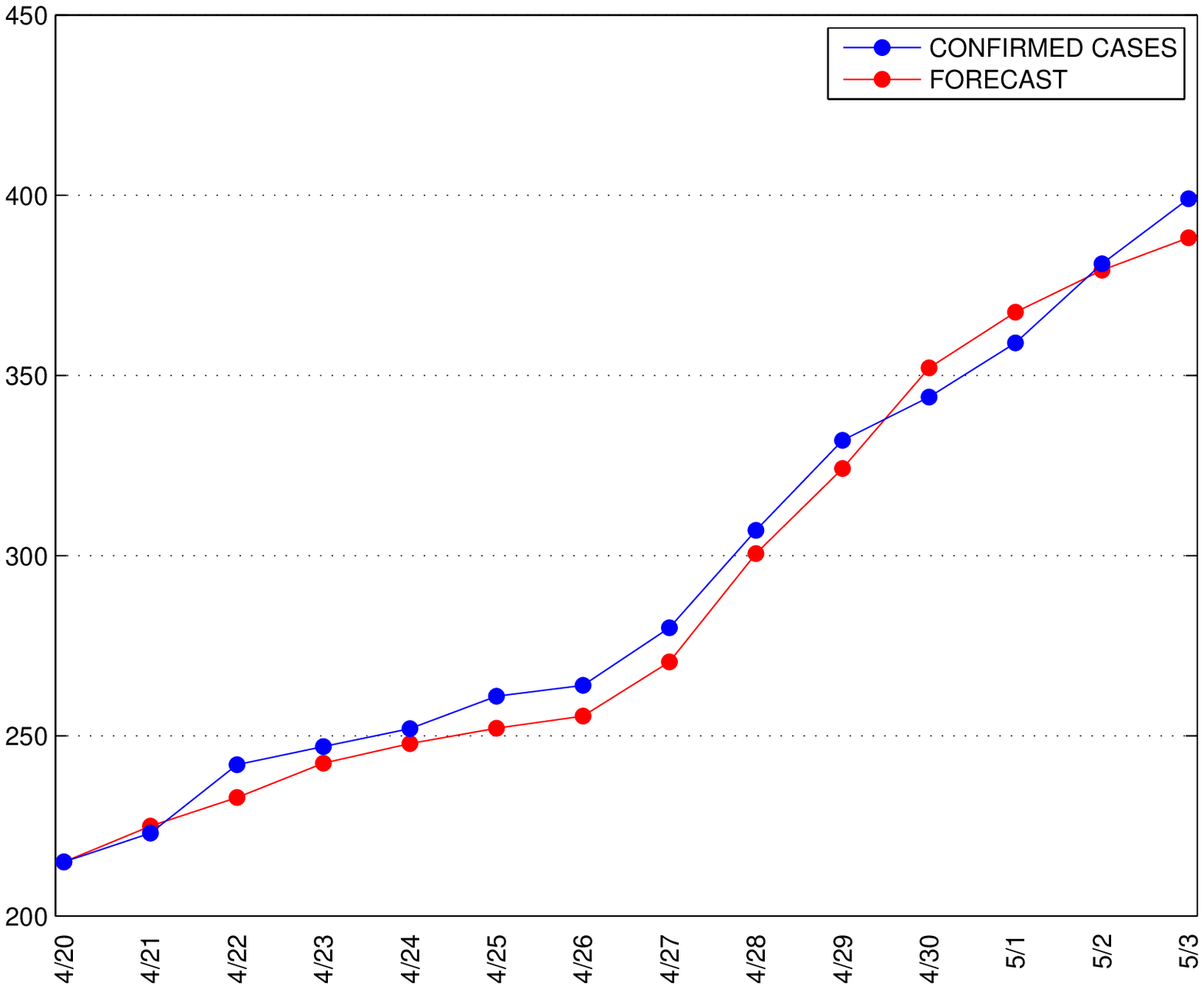} \\
\includegraphics[width=0.32\textwidth]{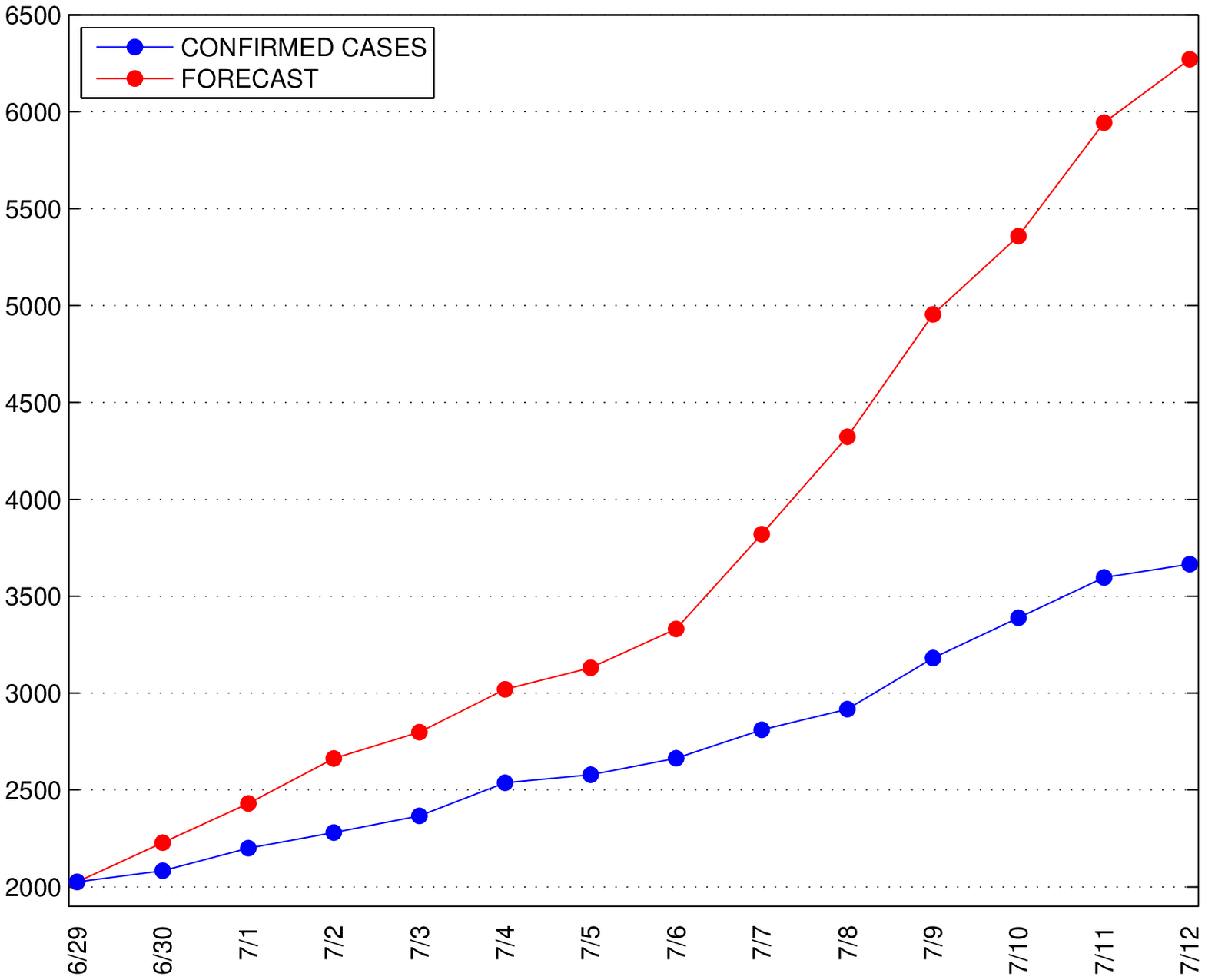} &
\includegraphics[width=0.32\textwidth]{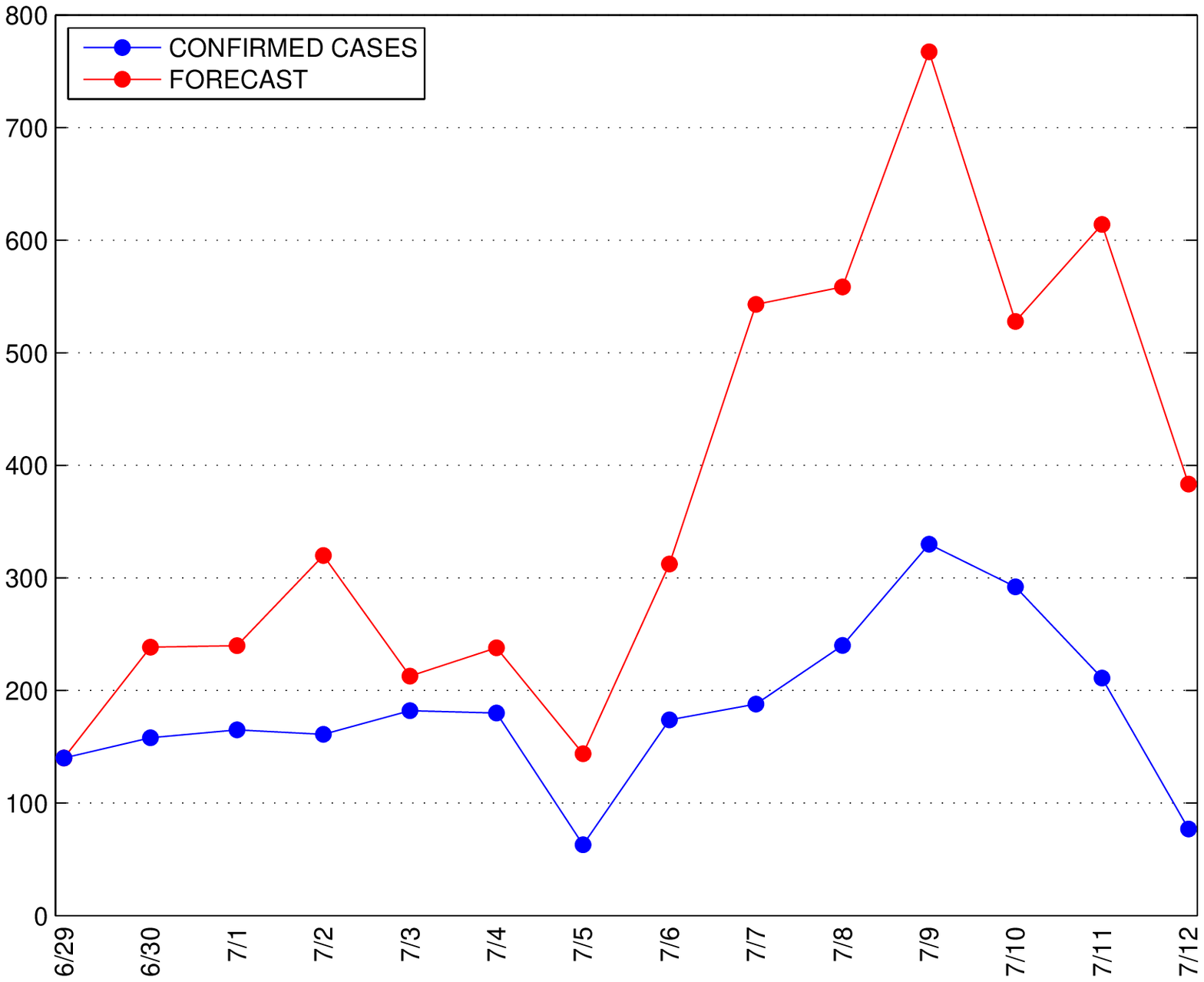} &
\includegraphics[width=0.32\textwidth]{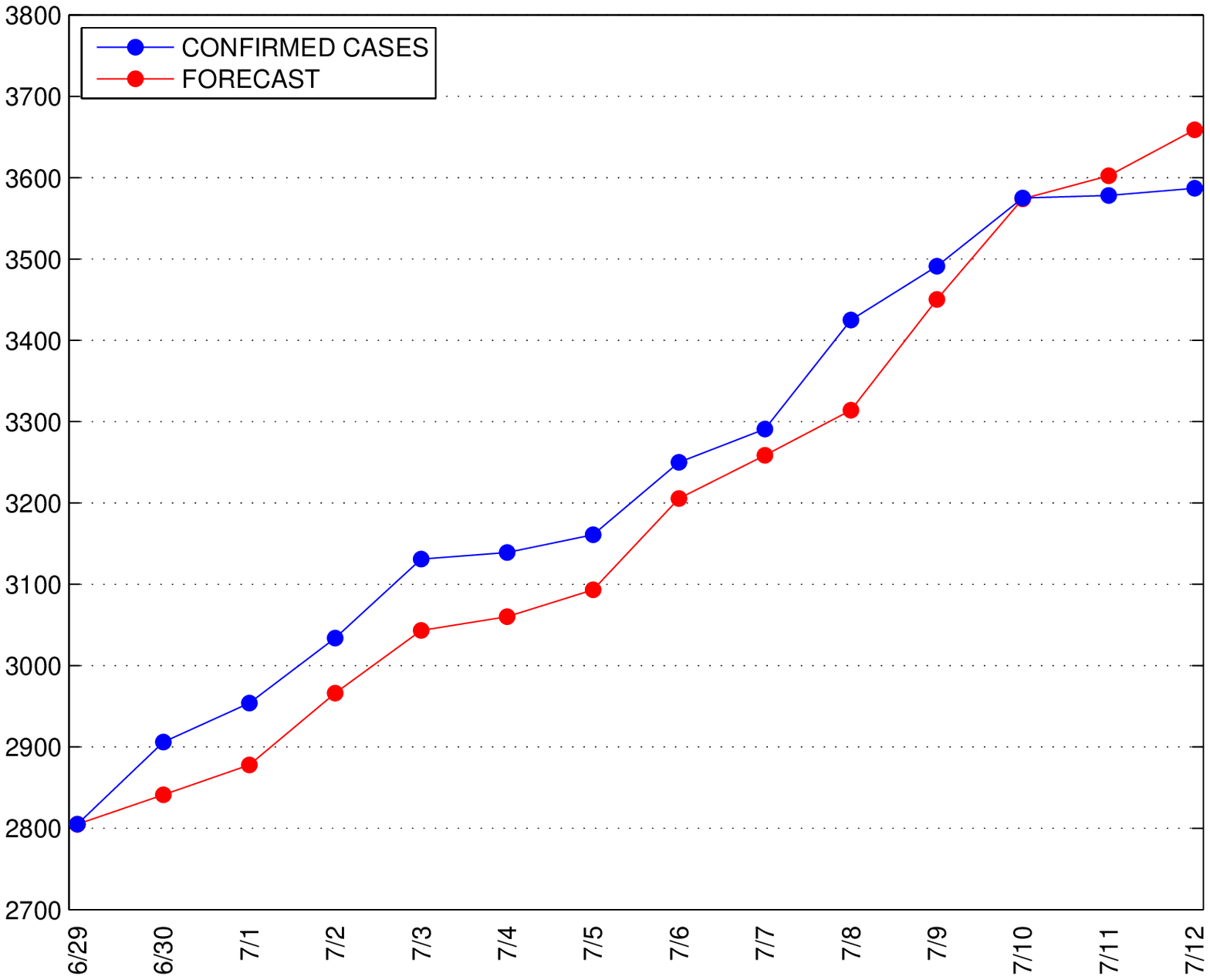} \\
\includegraphics[width=0.32\textwidth]{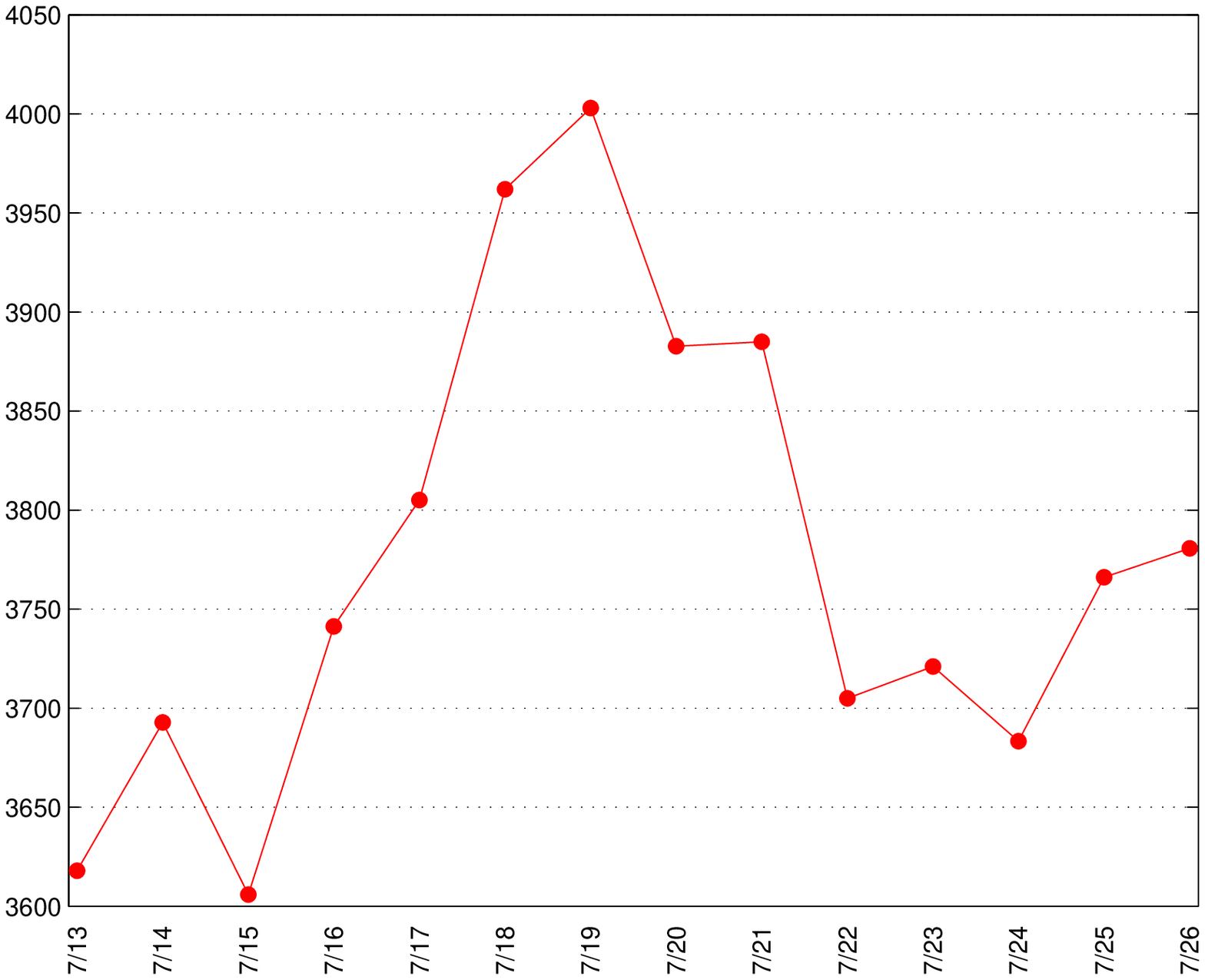} &
\includegraphics[width=0.32\textwidth]{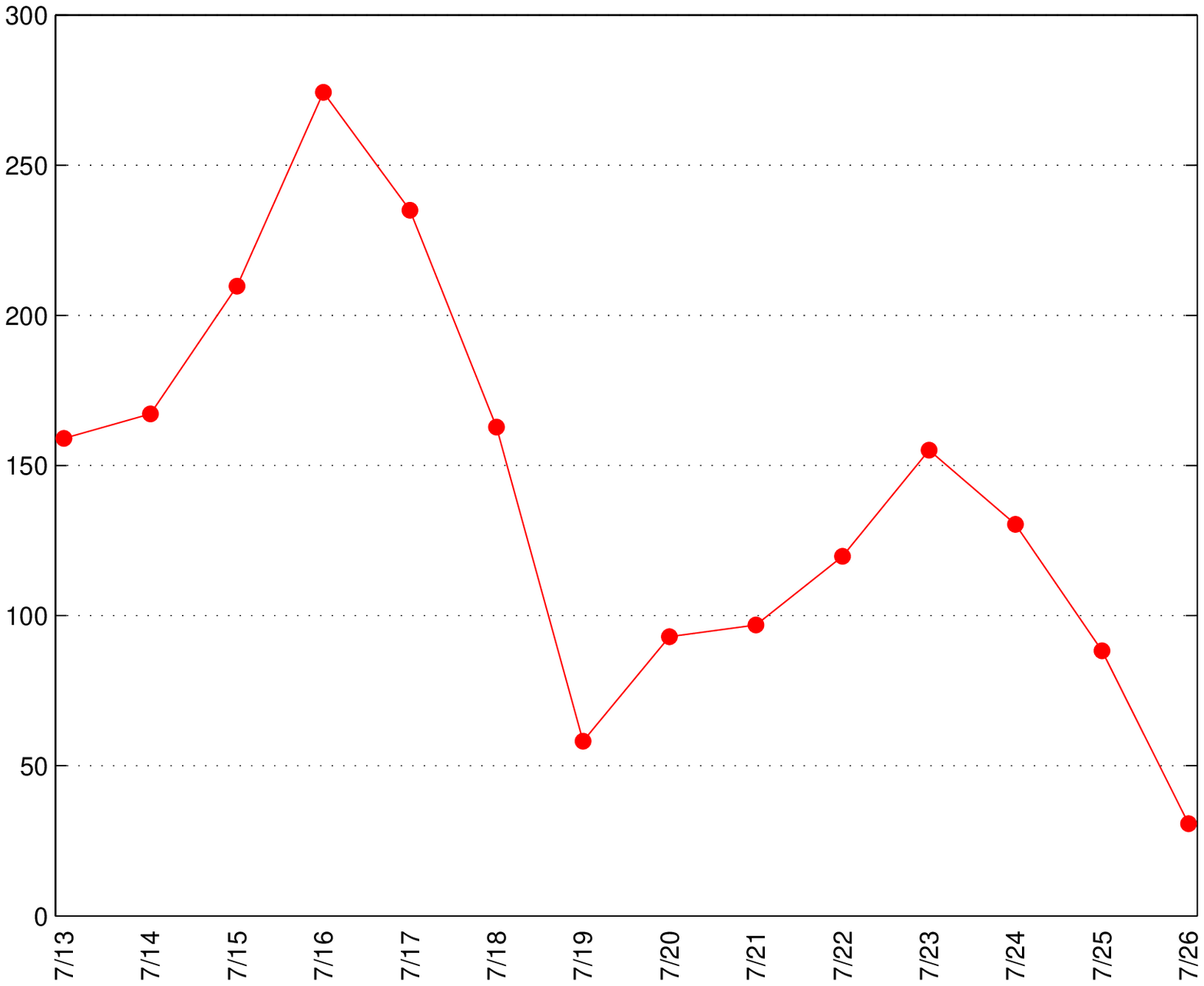} &
\includegraphics[width=0.32\textwidth]{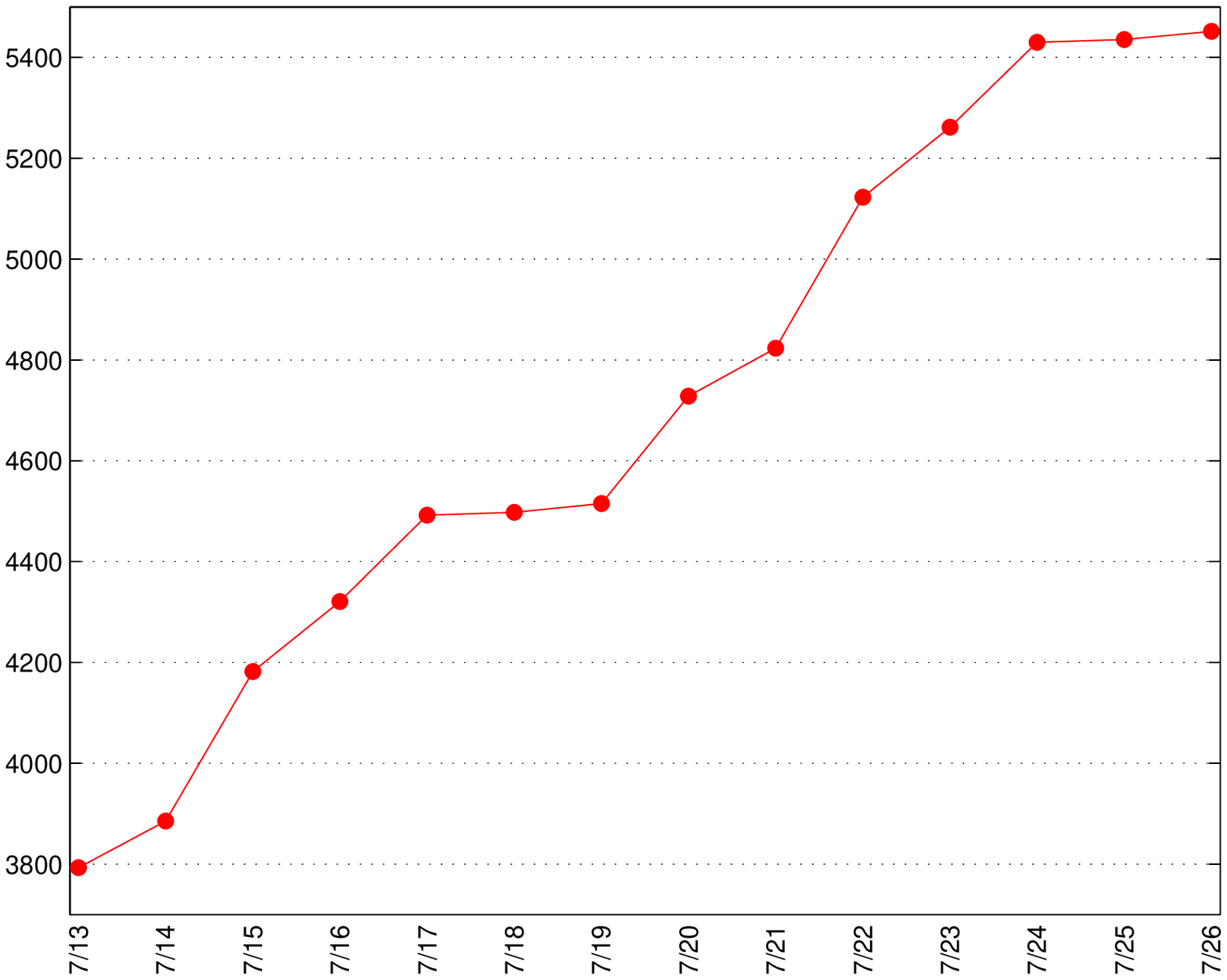} \\
\end{tabular}
\end{center}
\caption{The three forecasts. {\bf Row 1:} Complete data from the
beginning of the pandemia until today. {\bf Row 2:} First time
period: 4/20 -- 5/3. {\bf Row 3:} Second time period: 6/29 -- 7/12.
{\bf Row 4:} Third time period: 7/13 -- 7/26. Official data is in
blue, the predicted data is in red.} \vspace{-0.2cm}
\label{fig:forecasts}
\end{figure}

The experimental comparison between the official data and our
forecast  is illustrated on the second row of Figure
 \ref{fig:forecasts}. We observe good agreement between the predicted
and the real data. The number of actual cases (which is the biggest
value and the most unpredictable one) is well fitted by our
prediction. As expected, the longer the prediction period the larger
the error, and our first week prediction is overall more accurate
than the second week one. However, we managed to almost exactly
capture the daily cases numbers on days 8, 9 and 13, which is quite
a long-term forecast for the dynamics of the pandemic. The number of
recovered individuals (as the smallest one among the three) is
fitted the best.

Using  TD-SEIR model \eqref{eq:TDSEIR}  at the beginning of April
2020 we made forecast for COVID-19 spread among Bulgarian
population.
        Let us mention that on 12.04.2020 we sent to the Bulgarian National Operating Center (NOC)
     a forecast for an expected
    peak   the disease (of  new cases per day) in Bulgaria on 26.04.2020.
    Later (at the end of April) the Bulgarian NOC reported a peak in the period 20.04 --
    26.04. 2020.

For the second 14-days-time-frame June 29 -- July 12, based on the
data in Figure ~\ref{fig:beta_week} we predicted $\tb_{week_{i+1}}$
and $\tg_{week_{i+1}}$  via
$$
\tb_{week_{i+1}}=2\tb_{week_i}-\beta_{week_{i-1}},\qquad
\tg_{week_{i+1}}=\frac{\beta_{week_{i-5}}\tg_{week_i}+\beta_{week_{i-6}}\gamma_{week_{i-1}}}{2\beta_{week_{i-4}}}.
$$
Note that, at the starting day $t_K$ of the first and second
14-days-time-frames the weekly behavior of the infection rate
$\beta$ is quite similar. Indeed, we have had strong decrease in the
value between weeks $i-3$ and $i-2$, followed by an increase of size
at about half the decrease between weeks $i-2$ and $i-1$. During the
first weeks of prediction, the $\beta_{week}$ behavior again remains
similar: $\beta_{week_i}\approx \beta_{week_{i-3}}$ in both cases.
The big difference appears in the second week of prediction. It was
already mentioned about the strict government control measures
during the first time frame. On the contrary, there were very mild
government control measures during the second time frame -- Shopping
centers, parks, and even stadiums were open, high-school students
were taking final exams and were allowed to organize prom balls,
etc. As a result, the linear increase of $\beta_{week}$ holds true
up to date. Therefore, we used linear fitting for the infection rate
$\beta$ prediction. On the other hand, the percentage of healed
patients among the recovered individuals drastically increased,
meaning that simple averaging rule for the recovery rate $\gamma$
prediction was not adequate, any more. We conducted various
numerical experiments and concluded, that the weekly behavior of the
infection rate five weeks ago gives the most reliable information
about the recovery rate during the current week, i.e.,
$\beta_{week_{i-5}}\gamma_{week_i}\approx const$. Therefore, we
chose to fit this expression by a constant function. As input data,
we used the information from the previous two values, since
$\tg_{week_i}$ is again a prediction and not entirely reliable.
The experimental comparison between the official data and our
forecast for the second time frame is illustrated on the third row
of Figure \ref{fig:forecasts}. We observe an agreement between the
predicted and the real data during the first week of the time frame,
but not so much during the second one. As in the first forecast, the
number of predicted active cases is always larger than the
officially documented one, but the daily differences between the two
plots are significantly larger than before. Apart from modeling
obstructions, this behavior could also be explained with the
decreased percentage of critical cases and the decreased average age
of infected people. As a result, not all infected people test
themselves and the official data could be significantly lower than
reality. The same behavior is observed for the number of daily
cases. The number of recovered individuals is fitted well.

\begin{table}[ht]
\centering
\caption{Relative errors of the prediction.}\label{tab:errors}
 \begin{tabular}{|c|c c|c c|c c|c c|}
\hline
\multirow{3}{*}{Compartment} & \multicolumn{4}{|c|}{First time period}  & \multicolumn{4}{|c|}{Second time period}  \\ 
\cline{2-9} & \multicolumn{2}{|c|}{$\ell_2$} & \multicolumn{2}{|c|}{$\ell_\infty$} & \multicolumn{2}{|c|}{$\ell_2$} & \multicolumn{2}{|c|}{$\ell_\infty$} \\ \cline{2-9}
& Week 1 & Week 2 & Week 1 & Week 2 & Week 1 & Week 2 & Week 1 & Week 2\\\hline
Active cases & 0.026 & 0.052 & 0.036 & 0.083 & 0.159 & 0.564 & 0.214 & 0.711\\
Daily cases & 0.197 & 0.333 & 0.235 & 0.362 & 0.537 & 1.428 & 0.873 & 1.325\\
Recovered & 0.026 & 0.023 & 0.035 & 0.027 & 0.022 & 0.016 & 0.028 & 0.031\\
 \hline
\end{tabular}
\end{table}

In order to analyze the accuracy of the proposed prediction methodology, the relative weekly-based errors for the compartments ``Active cases'' $\{A_k\}_K^{K+13}$, ``Daily cases'' $\{D_k:=A_k-A_{k-1}\}_K^{K+13}$, and ``Recovered individuals'' $\{R_k\}_K^{K+13}$ for both the first and the second 14-days-time-frames are documented in Table~\ref{tab:errors}. Two different norms have been used - the $\ell_2$ one and the sup-norm $\ell_\infty$, i.e.,
$$
Err_A(\ell_2,week_i):=\frac{\|\widetilde A_{week_i}-A_{week_i}\|_2}{\|A_{week_i}\|_2},\qquad
Err_A(\ell_\infty,week_i):=\frac{\|\widetilde A_{week_i}-A_{week_i}\|_\infty}{\|A_{week_i}\|_\infty},
$$
for the active cases and analogously for the other compartments. We observe good agreement between the corresponding error margins of the two norms, which suggests robustness of the proposed model with respect to the norm choice within the $\{\ell_p\}$ family. Further, in agreement with Fig.~\ref{fig:forecasts}, we observe stable error behavior for the Recovered individuals, independent on the time period and the week of the period. For the remaining two compartments, we confirm that the second week prediction (days $\{t_k\}_{K+7}^{K+13}$) of each time period is less reliable than the first week prediction (days $\{t_k\}_{K}^{K+6}$) and that in the presence of strict government control measures our predictions are more accurate. Finally, we observe that the number of daily cases is the hardest one to predict, which is expected due to the lower values of that number and its day-by-day fluctuations.     

For the third 14-days-time-frame July 13 -- July 26 we apply the
same setup as for the second time frame, as they are consecutive
ones and nothing changed much in between. Here, we do not have
official data to compare with, since this time period is in the
future with respect to July 12. The forecast can be seen on the
forth row of Figure \ref{fig:forecasts}.

\section{Conclusions}\label{sec4}
In this paper, we applied a time-depended inverse SEIR model for
daily adjustment of the infection rate $\beta(t_k)$ and the recovery
rate $\gamma(t_k)$, based on a priori known information about the
number of susceptible individuals $S(t_k)$, the number of the active
cases $A(t_k)$, and the number of recovered individuals $R(t_k)$.
The problem becomes well-posed, if the latent rate $\omega(t_k)$ and
the number of infectious individuals $I(t_k)$ is also given. For the
former, we assume it to be a constant $\omega=1/T_e$, where the
average incubation period in Europe $T_e=4$ is used. For the latter,
we assume that $I(t_0)=A(t_0)$, i.e., all registered active cases at
day one of the pandemic were infectious, and further compute
$I(t_k)$, based on all the data from day $t_{k-1}$.

Next, we developed a difference scheme for solving the discretized
direct Cauchy problem that computes $S(t_k)$, $A(t_k)$, $I(t_k)$,
and $R(t_k)$, based on $\beta(t_{k-1})$, $\gamma(t_{k-1})$, and
$\omega(t_{k-1})$. Assuming, once again, that $\omega(t_k)$ is
constant, we developed a strategy for mid-time-term prediction of
all compartments of the host population. The strategy takes into
account the level of the government control measures, the new
cases/tests dependency on the day of the week, and the length of the
healing process. Three sets of numerical tests related to predicting
a 14-day-time-frame, based on the available official data for
Bulgaria up to the first day of the time frame were conducted. It
was observed that in all our experiments the number of recovered
individuals was well approximated along the whole time period. The
numbers of active and daily cases were well approximated during the
first week of the time period, but in the case of weak government
control measures and lack of social distancing, the predicted values
for the second week significantly exceeded the officially reported
ones. A potential explanation is that the truth is somewhere in
between, as there are infected people with mild or no symptoms, that
do not test themselves, thus, do not appear in the governmental
statistics. On the other hand, during strict government control
measures with compulsory social distancing, the numbers of active
and daily cases can be well approximated throughout the second week,
as well.

    \section{Acknowledgments}
The work of N. Popivanov and Ts. Hristov was partially supported by
the Sofia University under Grant 80-10-112/2020 and 80-10-122/2020.
The work of S. Margenov and S. Harizanov was partially supported by
the National Scientific Program "Information and Communication
Technologies for a Single Digital Market in Science, Education and
Security (ICTinSES)", contract No DO1–205/23.11.2018, financed by
the Ministry of Education and Science in Bulgaria.

\end{document}